\documentclass[a4paper,fleqn,usenatbib,useAMS]{mnras}

\usepackage{epsfig}
\usepackage{graphicx}
\usepackage{epstopdf}
\usepackage{morefloats}
\usepackage{amsmath}
\usepackage{natbib}
\usepackage{aas_macros}
\usepackage{comment}
\usepackage{subfigure}

\title[Be Discs in Misaligned Binaries]{Be discs in binary systems: II. Misaligned orbits}

\author[I. H. Cyr et al.]{I. H. Cyr$^{1}$, C. E. Jones$^{1}$, D. Panoglou$^{2}$, A. C. Carciofi$^{3}$, A. T. Okazaki$^{4}$
\\
$^{1}$Department of Physics and Astronomy, The University of Western Ontario, London, ON Canada N6A 3K7\\
$^{2}$Observat\'orio Nacional, Rua General Jos\'e Cristino 77, S\~ao Crist\'ov\~ao RJ-20921-400, Rio de Janeiro, Brazil\\
$^{3}$Instituto de Astronomia, Geof\'isica e Ci\^encias Atmosf\'ericas, Universidade de S\~ao Paulo, S\~ao Paulo, Brazil\\
$^{4}$Faculty of Engineering, Hokkai-Gakuen University, Toyohira-ku, 062-8605, Sapporo, Japan}

\begin{document}

\label{firstpage}
\pagerange{\pageref{firstpage}--\pageref{lastpage}}
\maketitle

\begin{abstract}
We use a smoothed particle hydrodynamics (SPH) code to examine the effects of misaligned binary companions on Be star discs. We systematically vary the degree of misalignment between the disc and the binary orbit, as well as the disc viscosity and orbital period to study their effects on the density in the inner and outer parts of the disc. We find that varying the degree of misalignment, the viscosity, and the orbital period affects both the truncation radius and the density structure of the outer disc, while the inner disc remains mostly unaffected. We also investigate the tilting of the disc in the innermost part of the disc and find the tilt increases with radius until reaching a maximum around 5 stellar radii. The direction of the line of nodes, with respect to the equator of the central star, is found to be offset compared to the orbital line of nodes, and to vary periodically in time, with a period of half a orbital phase. We also compare the scale height of our discs with the analytical scale height of an isothermal disc, which increases with radius as $r^{1.5}$. We find that this formula reproduces the scale height well for both aligned and misaligned systems but underestimates the scale height in regions of the disc where density enhancements develop.

\end{abstract}


\section{INTRODUCTION}
\label{intro}

The B-emission or Be stars are predominantly characterized by emission in the Balmer lines and have been studied since they were first detected by \citet{sec67}. In addition to the emission in the hydrogen lines, emission is often observed in singly or doubly ionized metals and is a direct result of radiative recombination in a disc-like distribution of gas. Other defining features include continuum excess in the visible up to the radio, linear polarization and variability over a significant range in period. These features originate in a geometrically thin disc of material ejected from the rapidly rotating central star. During the time since Secchi's discovery, our understanding of these star/disc systems has steadily improved but there are a significant number of remaining puzzles. The current status of Be-star research has been recently reviewed by \citet{riv13}.

In a pioneering paper, \citet{sha73} investigated the transfer of angular momentum in discs surrounding black holes and in doing so established a framework for future studies of hydrodynamic discs. The review paper by \citet{pri81} summarises the major properties of accretion discs including detailed discussion about the role of viscosity and emission processes. Since that time there have been a multitude of investigations that adopt and apply accretion theory to model astrophysical discs that surround a variety of different types of objects.
 
The same prescription originally developed for viscous accretion discs can be applied to study Be star discs except, in this case, the direction of the radial motion can switch between outward flowing, when the disc is being actively fed by the star, to inward flowing, as a pre-existing disc passively dissipates. In addition to the differences in the radial velocity, accretion discs fuel jets to shed angular momentum while decretion discs take the angular momentum that has been contributed from the central star and put material, at least some of it, into Keplerian orbit \citep{baa16}. The viscous decretion disc (VDD) scenario for Be stars, originally adopted by \citet{lee91} assumes that angular momentum is transferred from the star to the disc via an unknown mechanism. His work was followed up by \citet{oka01} who found that the outflow in the inner disc is subsonic and rotating in a Keplerian fashion, consistent with current findings about Be stars. The viscosity parameter is given by the usual $\alpha_\mathrm{SS}$ viscosity parameter and for Be stars is usually assumed to be in the range of 0.1 to 1 \citep[see][]{lee91}. The VDD has been the focus of an increasing number of studies.  For example, \citet{mar11} studied warping and precession of Be star discs and find that in non-coplanar binary systems the disc can become distorted and warped. \citet{fu15} studied oscillations, in particular the Kozai-Lidov mechanism, in hydrodynamic discs in binary systems by considering the effect of disc pressure and viscosity. They find that given sufficient time the disc becomes co-planar with the binary companion. \citet{lub15} find that in misaligned systems the disc can be more extended in radial distance than a disc in a co-planar system. Despite the fact the focus of \citet{fu15} and \citet{lub15} studies are about accretion discs, their work can help provide insight to the results presented in this investigation.

For the case of an isolated Be star, the central star supplies matter and angular momentum to the disc and is the primary energy input to the disc through its radiation field. The stellar wind potentially could ablate the disc material or help to constrain material to the equatorial regions especially in the inner parts of the disc \citep{oka12}. However, in a binary system the situation is more complex. In addition to tidally truncating the disc, the companion interacts with the disc through resonant torques which could cause the disc to warp and/or precess \citep{oka12, kee16} and these effects may complicate the density structure for the case of both aligned and misaligned discs.

The innermost disc is dominated by viscous torques so that isolated and binary systems may be similar in the inner regions \citep{oka12}. However, as explained by \citet{oka02}, with increasing radial distance the resonant torques due to the binary companion begin to play a bigger role. \citet{oka02} studied the gravitational effects of close companions on the structure of the discs in Be/X-ray binary systems and found that the disc was truncated, at a certain point, called the truncation radius, by the companion. Similar disc truncation have also been predicted in accretion systems with eccentric and misaligned binary companions by \citet{art94} and \citet{lar96}, respectively. This was also suggested by \citet{rei97} who studied Be/X-ray systems and found a relationship between the size of the H$\alpha$ emitting region and orbital period, demonstrating that the companion, in their case a neutron star, halted the flow of material and effectively truncated the disc. In a study of Be discs in coplanar binary systems, \citet{pan16} (hereafters Paper I) found that the position of the truncation radius depends on the parameters of both the disc (kinematic viscosity) and the binary (orbital period, mass ratio, eccentricity). 

Just inside the truncation radius, it was found that the disc density decreases at a slightly slower rate with increasing distance from the star than what is observed in discs around isolated Be stars \citep[][Paper I]{oka02}. As a result, the radial density distribution is flatter in binary systems and consequently discs in binary systems may be denser, and therefore more massive, than those in isolated systems. This phenomenon was called the ``accumulation effect". In Paper I, the effect was found to be stronger the smaller the viscosity and the orbital period and larger the mass ratio. Outside the truncation radius (i.e. the outermost portions of the disc), however, the density falls off at a substantially increasing rate. \citet{oka02} as well as Paper I also observed that $\alpha_\mathrm{SS}$ had an effect on the rate of the density fall-off in the outer part of the disc, with high viscosity discs having slower drop-off rates than low viscosity discs (see figure 11 of Paper~I). 

In addition to these changes in density, \citet{oka02} found that the surface density of the disc revealed the development of a two-armed spiral density wave at periastron \citep[figures 10 and 11 of][]{oka02} with one arm preceding the companion and the other located on the opposite side of the disc. Similar spiral waves were also observed in Paper~I.

Some progress has been made in recent years regarding the value of $\alpha_\mathrm{SS}$ in the disc. \citet{car12} modelled the 2003 dissipation phase of 28 CMa and found that $\alpha_\mathrm{SS}=1.0 \pm 0.2$. Subsequently, \citet{gho17} revisited this value to about 0.3, and extended the modelling to 4 complete cycles of disc formation and dissipations, spanning almost 4 decades of disc activity. They found that the values of $\alpha_\mathrm{SS}$ are systematically larger during disc build-up than at dissipation, and typically range between 0.1 and 1. In a recent work, \citet{rim17} modelled the lightcurves 54 Be stars in the SMC and found that the viscosity parameter is roughly two times larger at build-up ($\left\langle\alpha_{\rm SS}^{\rm bu}\right\rangle = 0.63$) than at dissipation ($\left\langle\alpha_{\rm SS}^{\rm d}\right\rangle = 0.29$)

In this study we use a numerical hydrodynamics code to expand on the works of \citet{oka02} and in Paper~I by investigating how these system evolve when the plane of the disc and binary are misaligned. Our goal is to understand how this misalignment, as well as various disc and orbital parameters, affect the density structure and dynamics of the disc. While much of the recent work in the literature is focused on X-ray and $\gamma$-ray Be binary systems, in this study we investigate, in a systematic fashion, the effect of a low mass, main sequence binary companion on Be star discs for misaligned orbits over a range of $\alpha_\mathrm{SS}$ and orbital period. The organization is as follows: Section 2 describes our methodology, results are presented in Section 3, and a discussion and summary is provided in Section 4.

\section{Methodology}

\subsection{Disc Modelling}

In order to study the hydrodynamical effects of a misaligned companion on the structure and dynamics of the disc, a three-dimensional (3D) smoothed particle hydrodynamics (SPH) code was used. This type of code solves the fluid equations by dividing the fluid into discrete elements called particles. The properties of each particle are smoothed over a finite spacial distance (smoothing length) using a kernel function in order to simulate a continuous fluid. The SPH code used in this work is based on the code developed by \citet{ben90a} and \citet{ben90b}. It was later modified by \citet{bat95} to include a second-order Runge-Kutta-Fehlberg integrator which uses individual time steps for each of the particles in order to decrease computational time. This code was later refined by \citet{oka02} for use in simulations of Be-binary systems. 

The disc systems constructed and followed in this work are binary systems where the primary is a Be star and the secondary is a low mass non-degenerate star assumed to be in a circular orbit. Each simulation starts as a disc-less system at the beginning of its disc building phase. The injection of mass into the disc is accomplished by placing a prescribed number of particles around the star at a radius of $r_{\mathrm{inj}} = 1.04 R_{\star}$, referred to as the injection radius. The position of these particles along $r_{\mathrm{inj}}$ is chosen randomly while their velocity is set to be Keplerian (no radial momentum). This is done for each time step of the simulation. All particles were given the same mass, which remained constant throughout the simulations. These masses, $m_{\mathrm{part}}$, are calculated based on the mass injection rate ($\dot{M}_{\mathrm{inj}}$) of the system and the number of particles ($N_{\mathrm{part}}$) created during each mass injection event, both of which are constant throughout the simulations. For this work these values were set to $\dot{M}_{\mathrm{inj}}= 10^{-8}$ M$_{\sun}$/yr and $N_{\mathrm{part}}=6000$, resulting in $m_{\mathrm{part}} \approx 4\times10^{-15}$ M$_{\sun}$.

Once injected into the disc, particles interact with each other through viscosity, which transfers angular momentum throughout the disc. Most particles eventually fall back onto the star while the remainder move outward, carrying angular momentum with them away from the central star. As shown by \citet{rim17}, the fraction that falls back is about $\dot{M}_{\rm inj} [2.-(R_{\mathrm{inj}}/R_{\mathrm{star}})^{1/2}]$, or about 98$\%$ in our case. The Shakura-Sunyaev viscosity $\alpha_\mathrm{SS}$-prescription \citep{sha73} is used to define the viscosity of the system. Details about the implementation of the viscosity into the SPH code can be found in \citet{oka02}. The disc is assumed to be isothermal, therefore the gas particles are all set to the same temperature. The disc temperature ($T_{\mathrm{disc}}$) was set to 60\% of the effective temperature of the star, which was found to be a good approximation for the temperature structure of the disc \citep{car06,mil98}. 

The stars are modelled using sink particles delimited by an accretion radius inside which any particle is assumed to have accreted onto the star and is therefore removed from the simulation. The accretion radius of the primary (Be star) is equal to the radius of the star, which in this work was set to 3.67 $R_{\star}$, typical of a main-sequence B5 star. The accretion radius for the secondary is determined using the Eggleton approximation of the Roche lobe $R_{\mathrm{L}}$.

\begin{equation}
R_{\mathrm{L}} = \frac{0.49 q^{2/3}}{0.6 q^{2/3}+\ln(1+q^{1/3})} a,
\label{eq:fit_rt}
\end{equation}
where \textit{q} is the mass ratio of the stars ($q = M_{\mathrm{sec}}$/$M_{\mathrm{prim}} < 1$) and \textit{a} is the separation distance between them. Note that \textit{a} changes over time for eccentric systems, as is discussed in Paper I, however since we focus on circular orbits in this work $R_{\mathrm{L}}$ remains fixed throughout each simulation. The masses, radii, and effective temperatures of the stars were based on the parameters of the Pleione binary system \citep{har88,hir07,nem10}. It's important to note that this study does not focus on the modelling of this particular binary system, but instead aims to study the general behaviour of Be discs in close binary systems. For this purpose, we have modified a few parameters, such as the orbital period and eccentricity of the system in order to reduce the complexity of the simulations. All these parameters can be found in Table~\ref{table:parameters}.

In order to investigate the effect of the companion on the disc, we varied three parameters: the misalignment angle $\theta$, defined as the angle between the stellar equator of the primary and the plane of the binary orbit, the viscosity parameter, $\alpha_\mathrm{SS}$, and the orbital period of the binary. Simulations were run for four misalignment angles ($\theta$ = 0$\degr$, 30$\degr$, 45$\degr$, 60$\degr$), three values of viscosity ($\alpha_\mathrm{SS}$ = 0.1, 0.5, 1.0), and for binary systems with short (30 days) and long (60 days) orbital periods, corresponding to a semi-major axis of 59.9 $R_{\sun}$ and 95.1 $R_{\sun}$, respectively.

\begin{figure}
\includegraphics[width=\columnwidth]{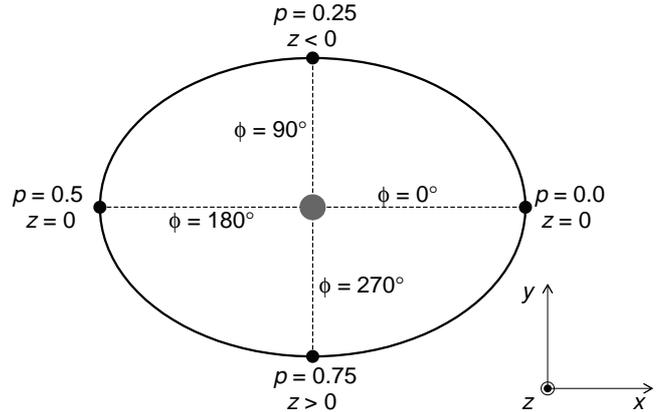}
\caption{Graphical representation of a misaligned system from a top-down view perpendicular to the plane of the disc. The gray circle represents the Be star (primary) while the smaller black circles show the positions of the secondary at various orbital phases, $p$, along a projection of its orbital path (solid line). Also shown is whether the secondary is above ($z>0$), below ($z<0$), or in ($z=0$) the plane of the disc. The dashed lines show the direction of the four principal azimuthal angles $\phi$ in relation to the phases of the secondary. The axes are defined such that the \textit{x}-axis points to the right, the \textit{y}-axis upward, and the \textit{z}-axis points out of the page, as indicated at the bottom right of the Figure.}
\label{fig:orbit}
\end{figure}

Following Paper I, we define $p$ as the orbital phase of the secondary during a single orbit, with $0 \leq p < 1$. We examine four orbital phases, $p$ = 0.00, 0.25, 0.50, and 0.75. Figure~\ref{fig:orbit} shows a graphical depiction of this notation scheme. Here the $x-y$ plane is defined as the plane of the stellar equator with the $z$ axis pointing out of the figure. The Be star is represented by the gray circle at the center of the figure while the solid black line shows the path of the secondary\footnote{Note that this is the orbital path projected onto the plane of the disc, explaining why the orbit appears elliptical.}. For each position of the secondary (black circles) we have indicated the corresponding orbital phase $p$ and whether the secondary is above ($z>0$), below ($z<0$), or in ($z=0$) the plane of the disc. Note that in the aligned case ($\theta = 0$) the path of the secondary is circular and the secondary stays in the plane of the disc at all times. A cylindrical coordinate system ($r$,$\phi$,$z$) is used when describing specific locations or cross-sections inside the disc, with the origin defined at the primary and the ($r$,$\phi$,0)-plane defined as the plane of the stellar equator. The angle $\phi$ is defined from the positive $x$-axis moving counter-clockwise. The dashed lines in Figure~\ref{fig:orbit} show the directions of the azimuthal angles $\phi$ = 0$\degr$, 90$\degr$, 180$\degr$, and 270$\degr$.

\begin{table}
 \caption{Adopted parameters of the binary system.}
 \label{table:parameters}
 \begin{tabular}{lc}
  \hline
  Parameters & Values\\
  \hline
  Mass of primary & 2.9 M$_{\sun}$\\ 
  Radius of primary & 3.67 R$_{\sun}$\\ 
  Effective temperature & 12000 K\\
  \hline
  Mass of secondary & 0.31 M$_{\sun}$\\
  Radius of secondary & 0.38 R$_{\sun}$\\
  \hline
  Eccentricity & 0\\ 
  Mass injection rate & 10$^{-8}$ M$_{\sun}$/yr\\
  Particles mass & $4\times10^{-15}M_{\sun}$\\
\hline
  Viscosity parameters & 0.1 / 0.5 / 1.0\\
  Misalignment angles & 0$\degr$ / 30$\degr$ / 45$\degr$ / 60$\degr$\\
  Period (days) & 30 / 60s\\
\hline
 \end{tabular}
\end{table}

\section{Results}

\subsection{Average Surface Density}
\label{sec:avrg_sigma}
It is informative to calculate some of the average features of the disc prior to investigating localized changes or perturbations. We start with an investigation of the average disc density. This is accomplished by transferring the particles to a cylindrical grid centred at the primary. The density is then integrated vertically to obtain the azimuthally averaged surface density ($\left < \Sigma(r)\right > _{\phi}$). This procedure was adopted for Paper~I.

One important aspect we first consider is whether our simulations are sufficiently developed so that the only major variations are those caused by the motion of the secondary. As \citet{hau12} showed, discs around isolated Be stars will continue to grow as long as there is a continuous and constant injection of mass and angular momentum from the star, meaning they will never reach a state of perfect equilibrium. However we also know that the growth of the disc, although fast in the beginning of its building phase, slows dramatically as the disc expands outward \citep{hau12}. The discs should therefore reach a quasi-static state (QSS), where the disc growth become insignificantly small within the span of one orbital period. It is also important to note that, since Be star discs are built from the inside out, the inner parts reaches QSS much faster than the outer parts. See section 2.1 of Paper~I for a more in depth discussion.

\begin{figure}
\subfigure[30 day period, $\alpha_\mathrm{SS}=0.1$]{\includegraphics[width=0.49\columnwidth]{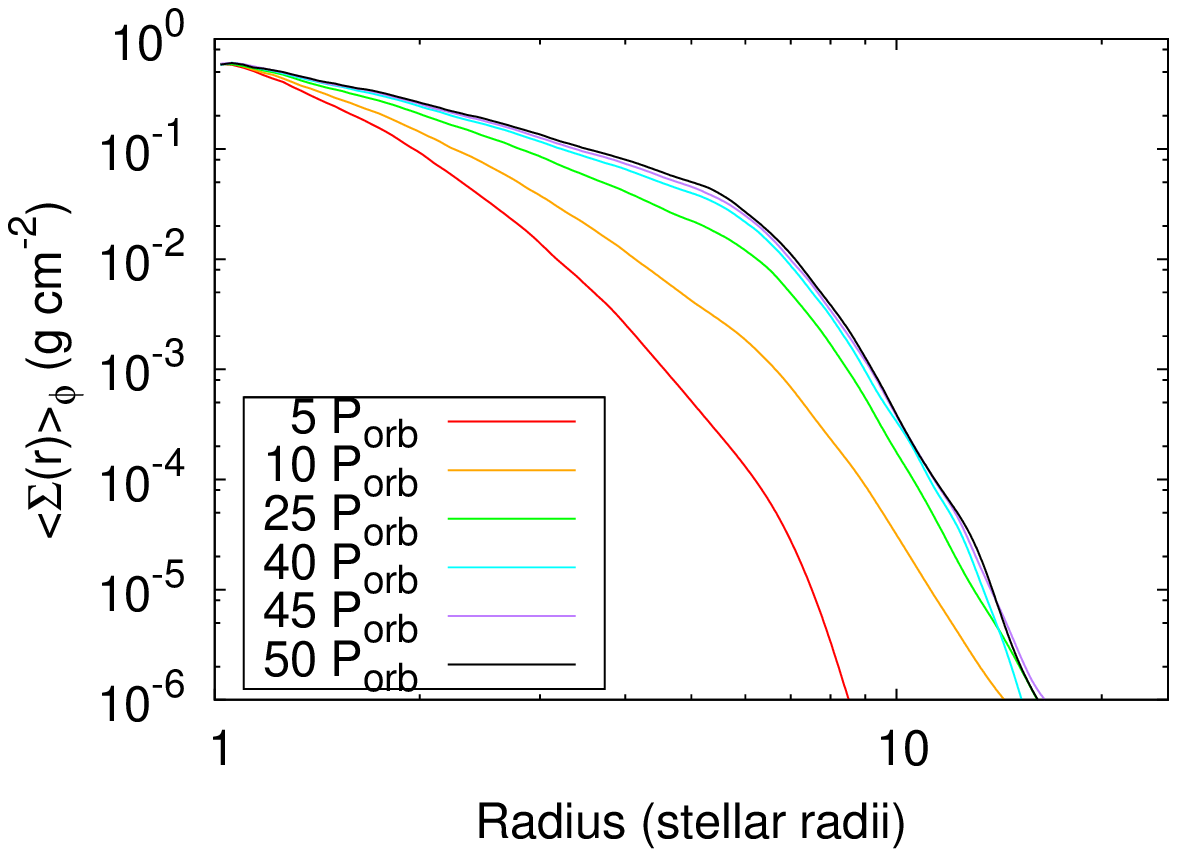}} 
\subfigure[60 day period, $\alpha_\mathrm{SS}=0.1$]{\includegraphics[width=0.49\columnwidth]{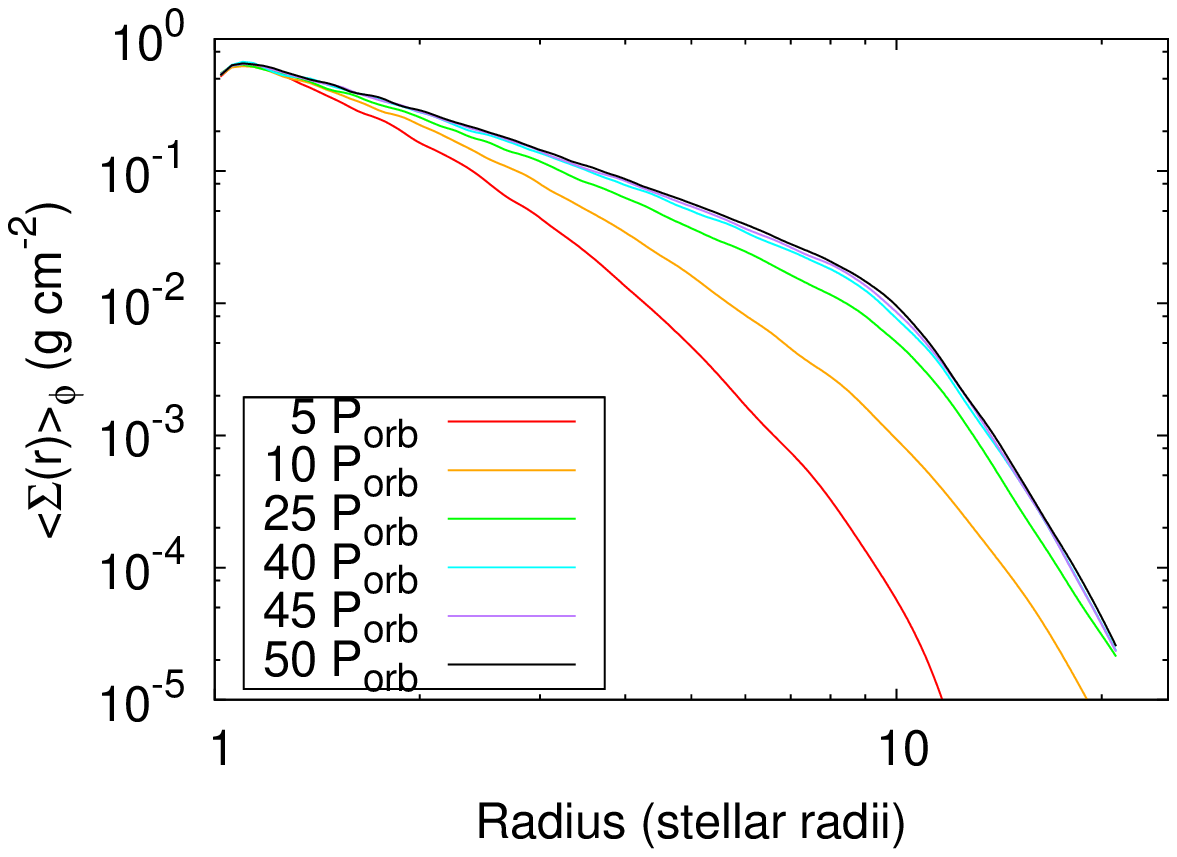}} \\
\subfigure[30 day period, $\alpha_\mathrm{SS}=0.5$]{\includegraphics[width=0.49\columnwidth]{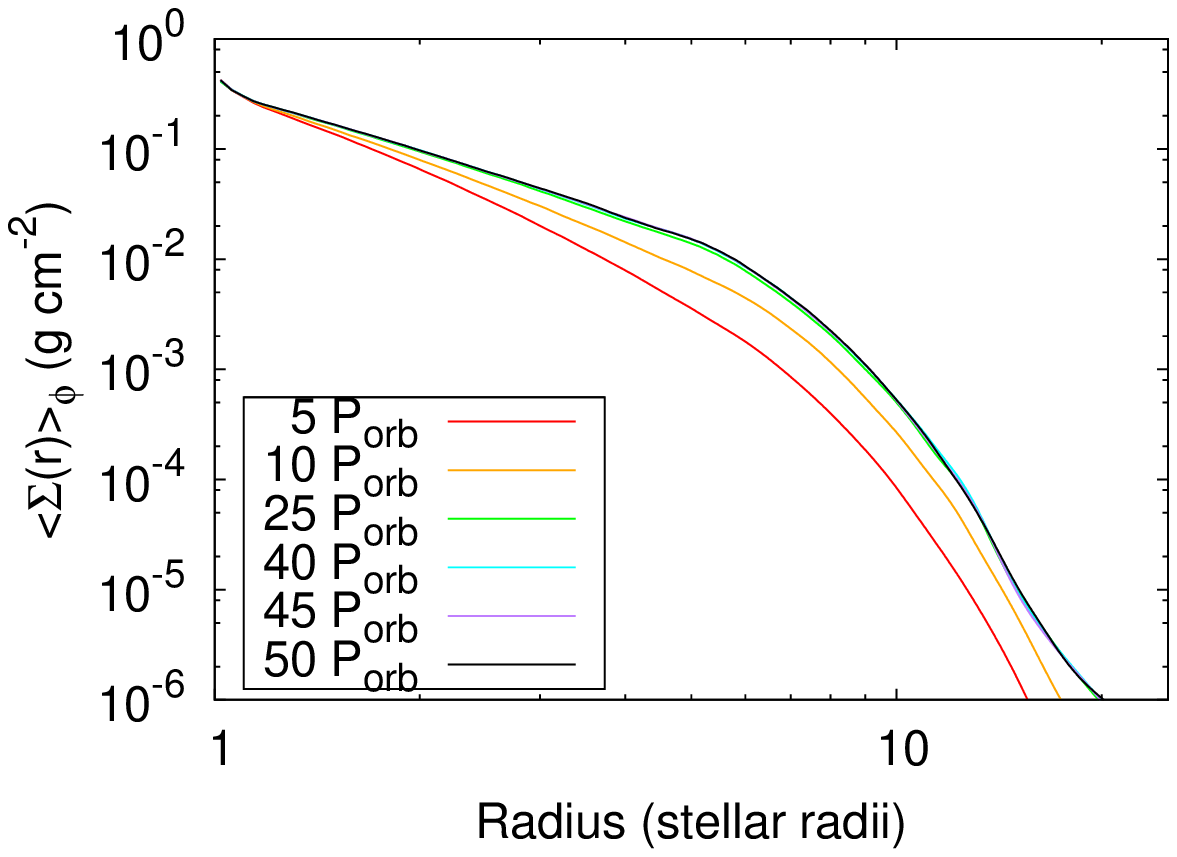}} 
\subfigure[60 day period, $\alpha_\mathrm{SS}=0.5$]{\includegraphics[width=0.49\columnwidth]{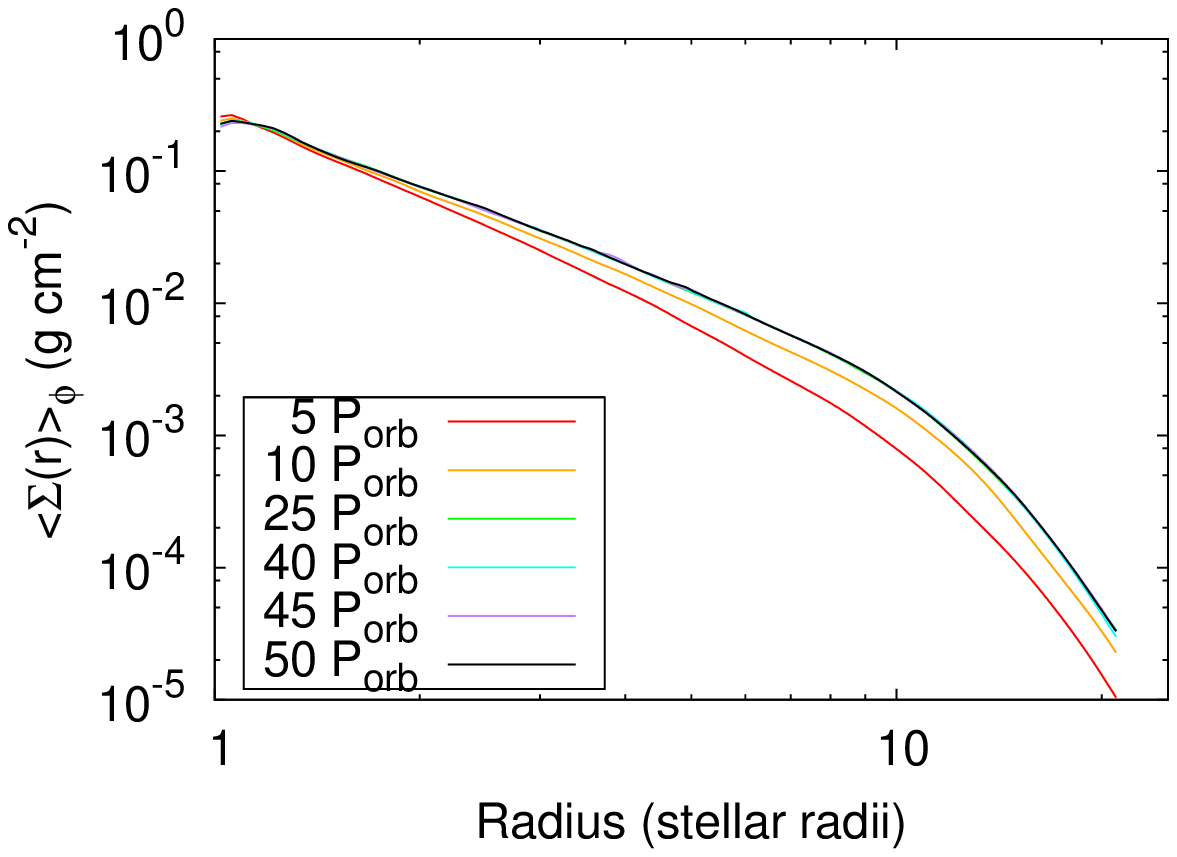}} \\
\subfigure[30 day period, $\alpha_\mathrm{SS}=1.0$]{\includegraphics[width=0.49\columnwidth]{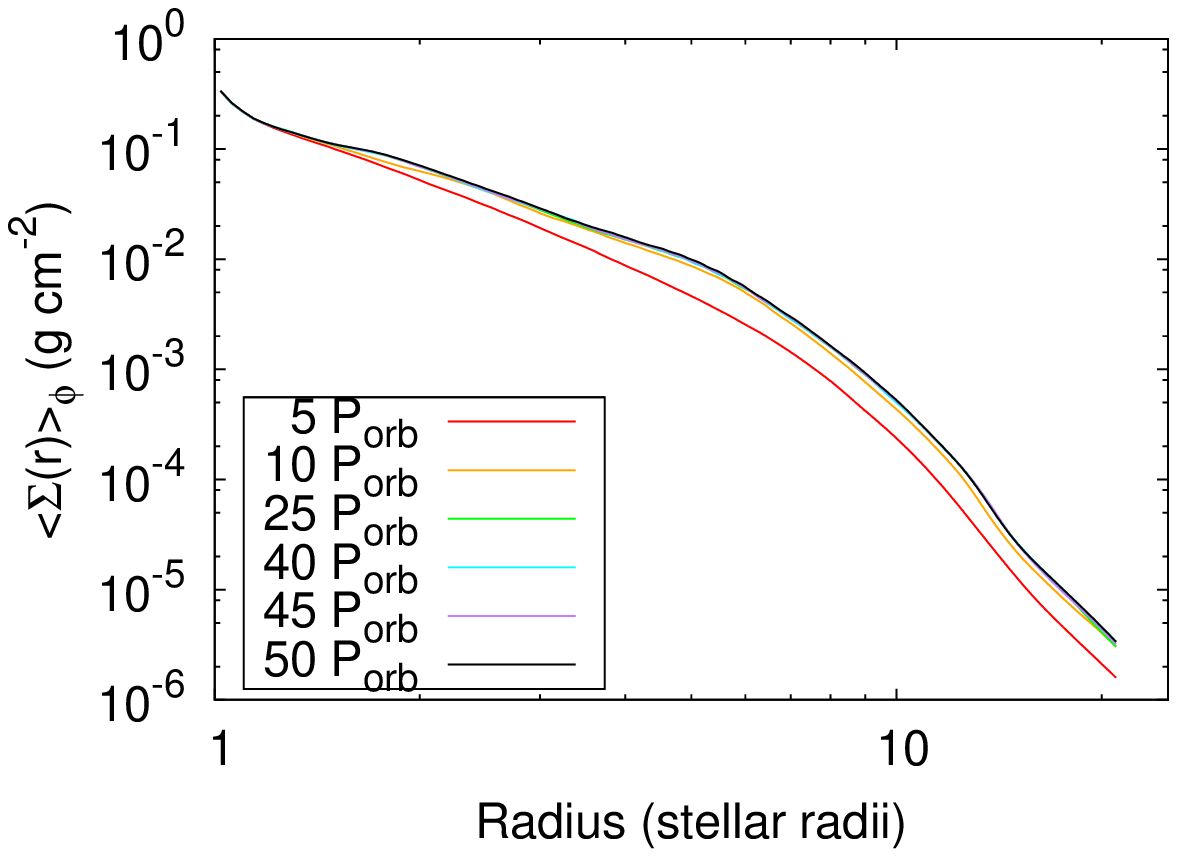}} 
\subfigure[60 day period, $\alpha_\mathrm{SS}=1.0$]{\includegraphics[width=0.49\columnwidth]{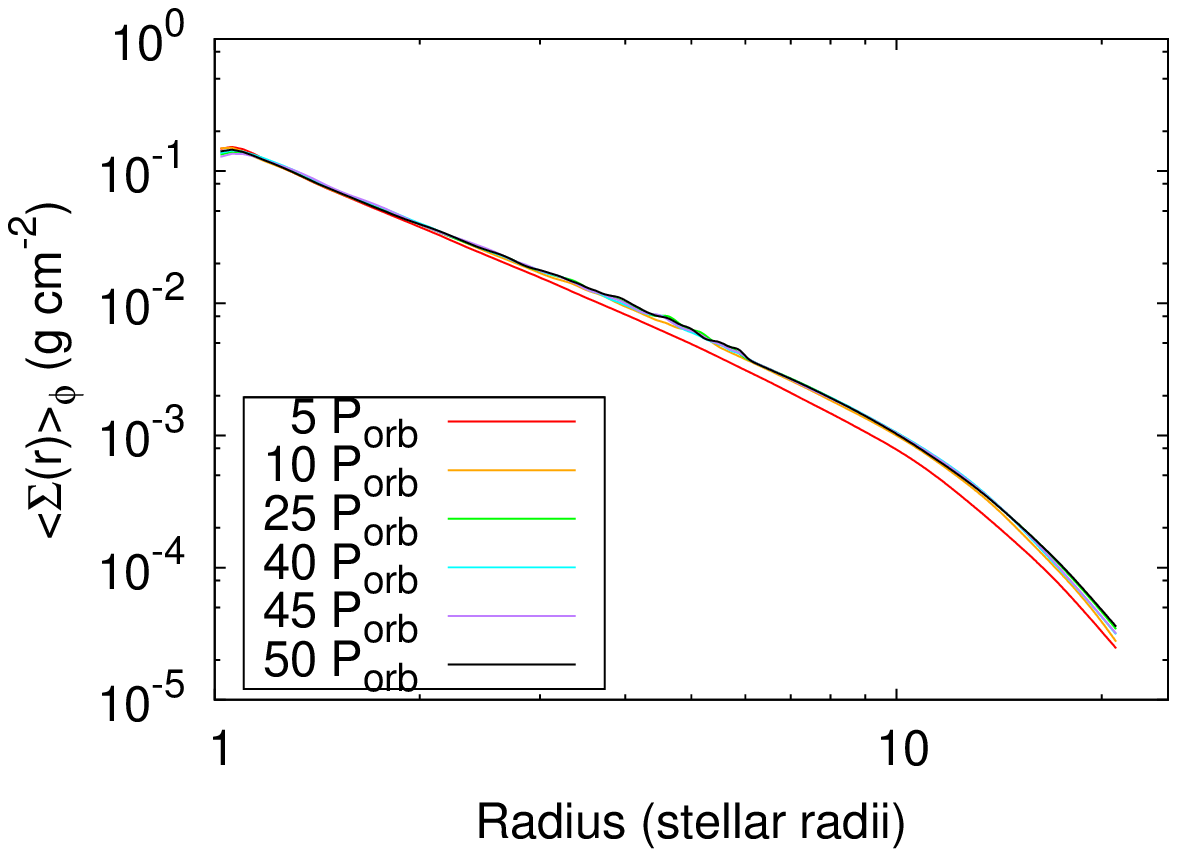}} 
\caption{Temporal evolution of the azimuthally averaged surface density, $\left < \Sigma(r)\right > _{\phi}$, as a function of radius for the short (left column) and long (right column) orbital period systems, and for $\alpha_\mathrm{SS}$ values of 0.1 (top row), 0.5 (middle row), and 1.0 (bottom row). All simulations have a misalignment angle of 0$\degr$}
\label{fig:qss}
\end{figure}

Figures~\ref{fig:qss}(a) through (f) show the temporal evolution of the azimuthally averaged surface density, $\left < \Sigma(r)\right > _{\phi}$, at various times during the disc building phase (see legend). The leftmost and rightmost panels show the results for short and long period systems, respectively, with each row, from top to bottom, for $\alpha_\mathrm{SS}$ of 0.1, 0.5, and 1.0, respectively. Each simulation was run up to 50 orbital periods ($P_{orb}$). Note that only the aligned systems ($\theta = 0$) are shown in Figure~\ref{fig:qss}, however we have found the QSS time is virtually the same for the corresponding misaligned cases.

We see that all simulations have reached QSS by 50 $P_{orb}$, although the higher viscosity discs reach their QSS much faster than lower viscosity discs. This result is similar to the temporal evolution of the surface density profile found in Paper I for aligned systems and by \citet{hau12} for isolated Be stars; both found that discs with higher viscosity reach their steady-state faster than those with low viscosity. This happens because the $\alpha_\mathrm{SS}$ acts as a timescale factor in the hydrodynamical equations governing theses discs \citep{pri81}. The number of active particles for the short period systems after 50 $P_{orb}$ is presented in Table~\ref{table:num_particle}.

\begin{table}
 \caption{Approximate number of active particles after 50 $P_{orb}$ for short period (30 day) systems.}
 \label{table:num_particle}
 \begin{tabular}{ccc}
  \hline
  Viscosity & Misalignment  & Number of active\\
  parameter & angle  & particles\\
  \hline
  & 0$\degr$ & 100,000\\
$\alpha_\mathrm{SS} = 0.1$  & 30$\degr$ & 79,000 \\
 & 45$\degr$ & 75,000 \\
 & 60$\degr$ & 18,000 \\
  \hline
  & 0$\degr$ & 37,000\\
$\alpha_\mathrm{SS} = 0.5$  & 30$\degr$ & 36,000\\
 & 45$\degr$ & 37,000\\
 & 60$\degr$ & 39,000\\
  \hline
  & 0$\degr$ & 25,000\\
$\alpha_\mathrm{SS} = 1.0$ & 30$\degr$ & 25,000\\
 & 45$\degr$ & 25,000\\
 & 60$\degr$ & 26,000\\
\hline
 \end{tabular}
\end{table}

\begin{figure}
\subfigure[30 day period, $\alpha_\mathrm{SS}=0.1$]{\includegraphics[width = 0.49\columnwidth]{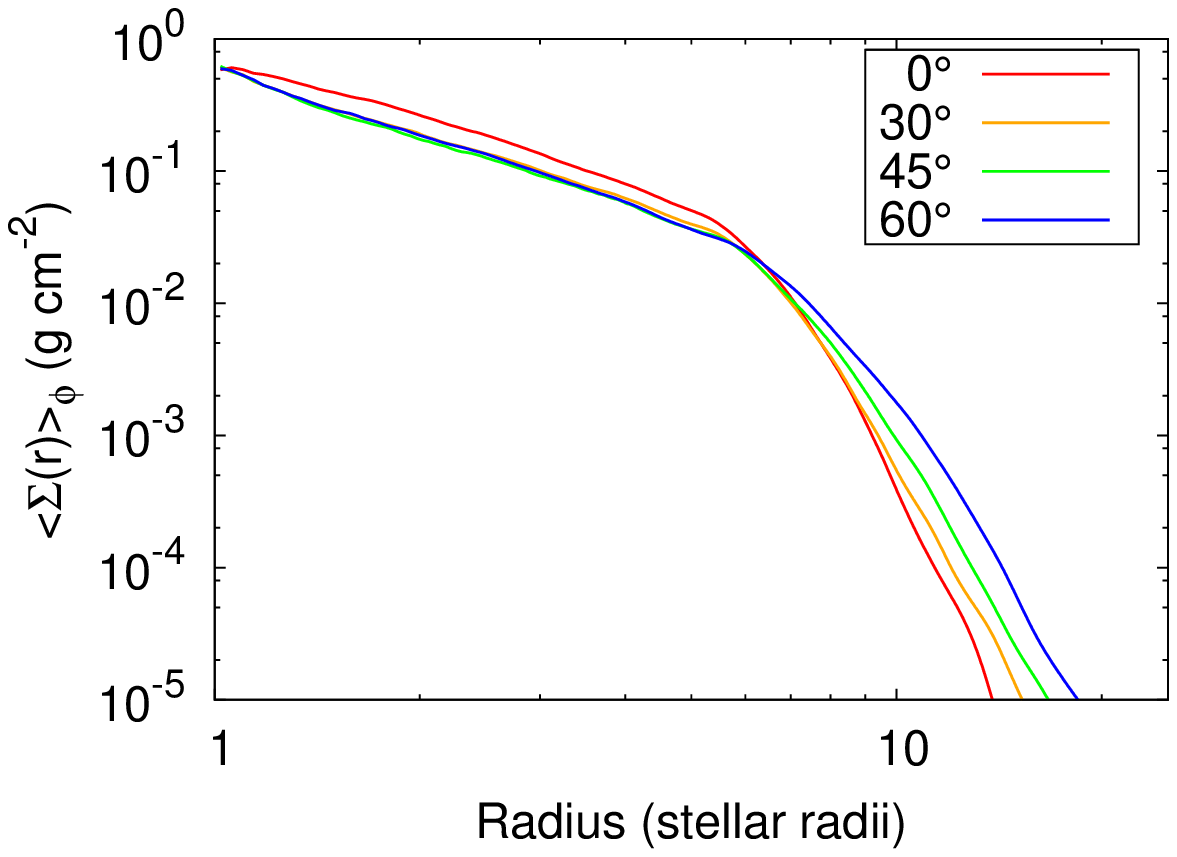}} 
\subfigure[60 day period, $\alpha_\mathrm{SS}=0.1$]{\includegraphics[width = 0.49\columnwidth]{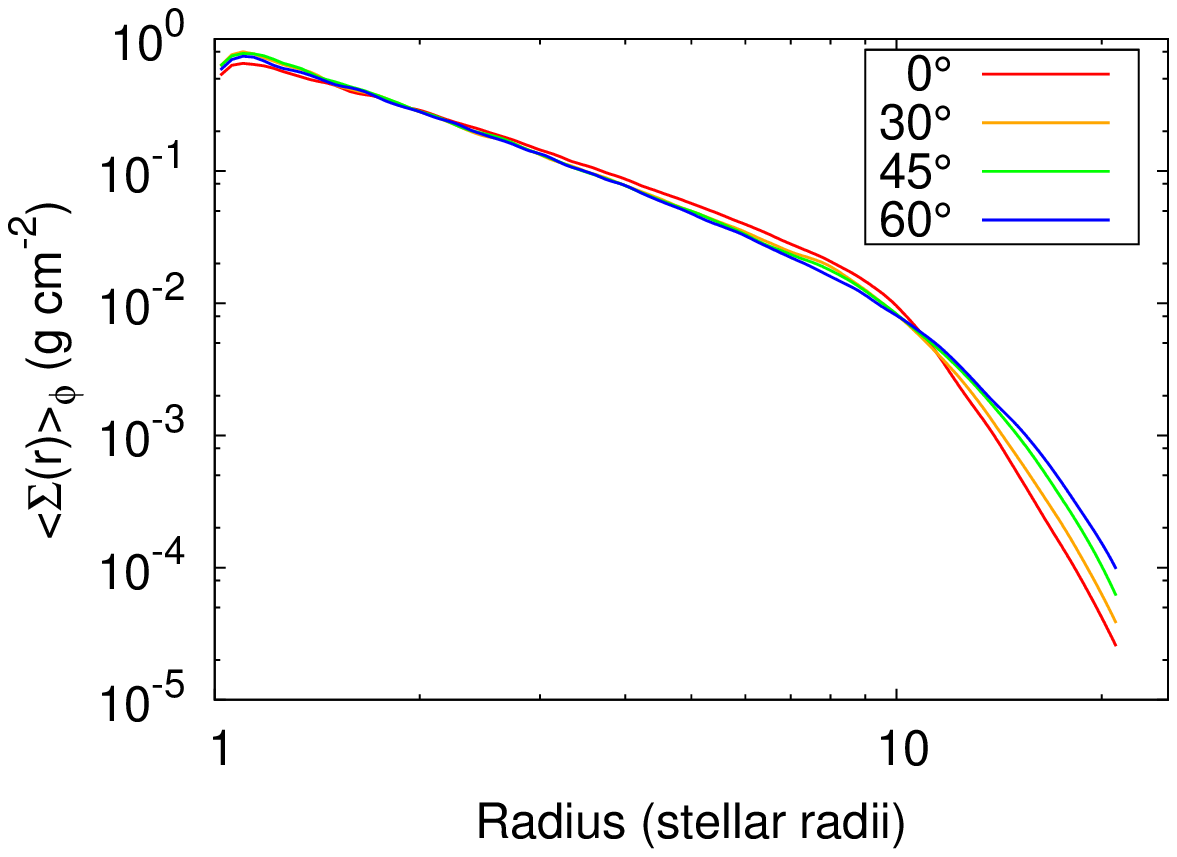}} \\
\subfigure[30 day period, $\alpha_\mathrm{SS}=0.5$]{\includegraphics[width = 0.49\columnwidth]{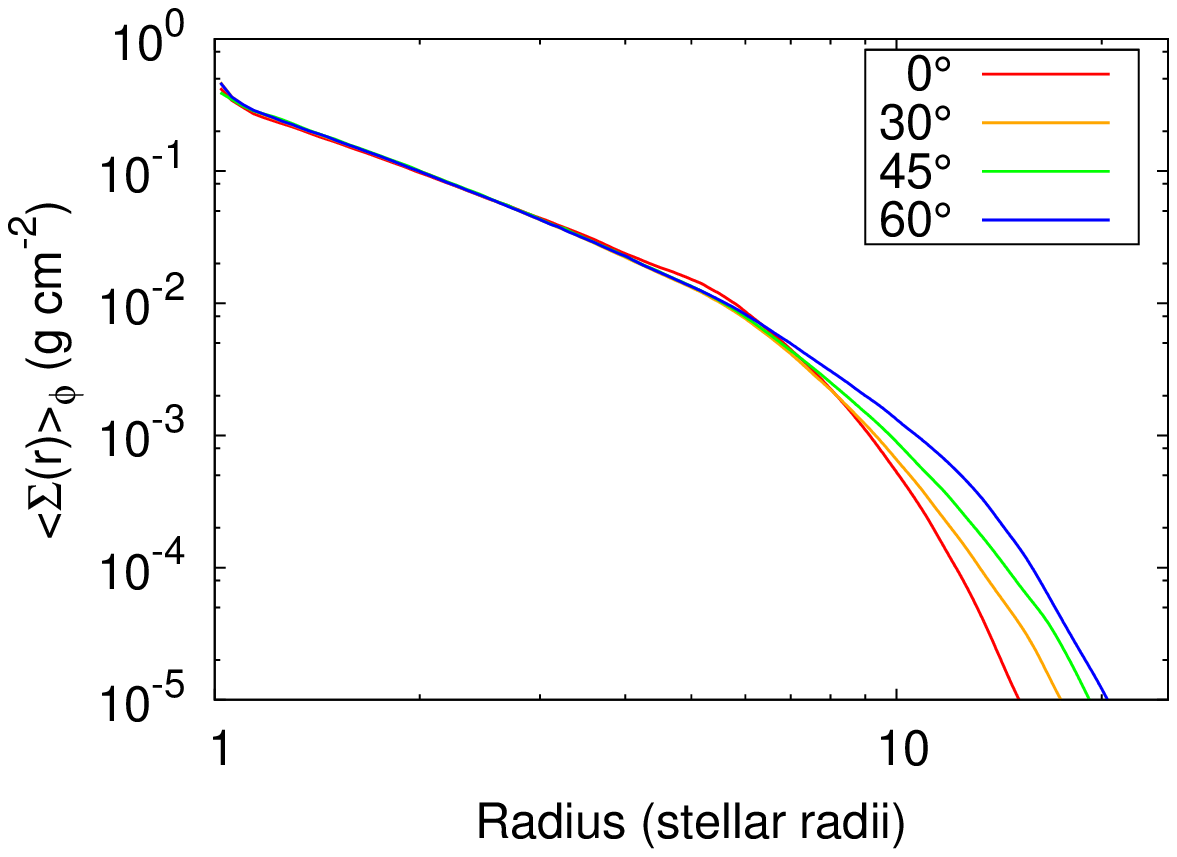}} 
\subfigure[60 day period, $\alpha_\mathrm{SS}=0.5$]{\includegraphics[width = 0.49\columnwidth]{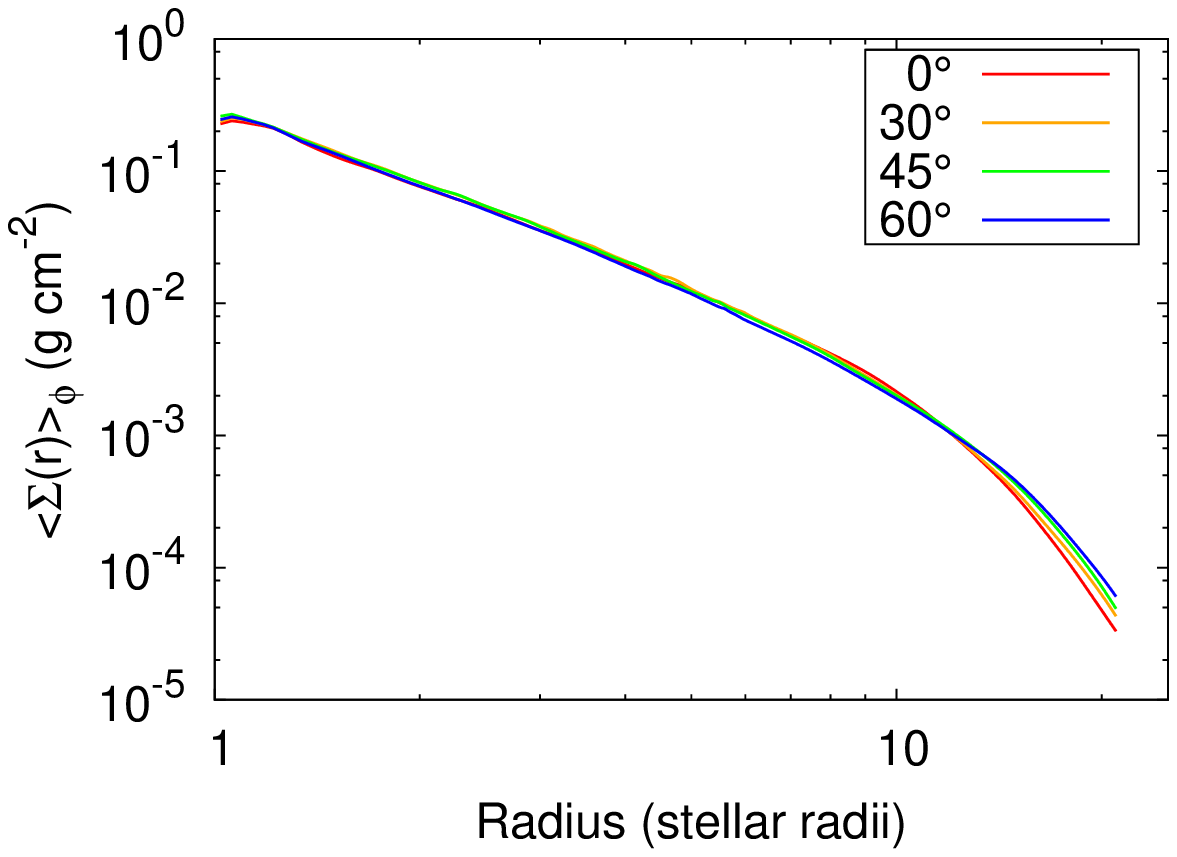}} \\
\subfigure[30 day period, $\alpha_\mathrm{SS}=1.0$]{\includegraphics[width = 0.49\columnwidth]{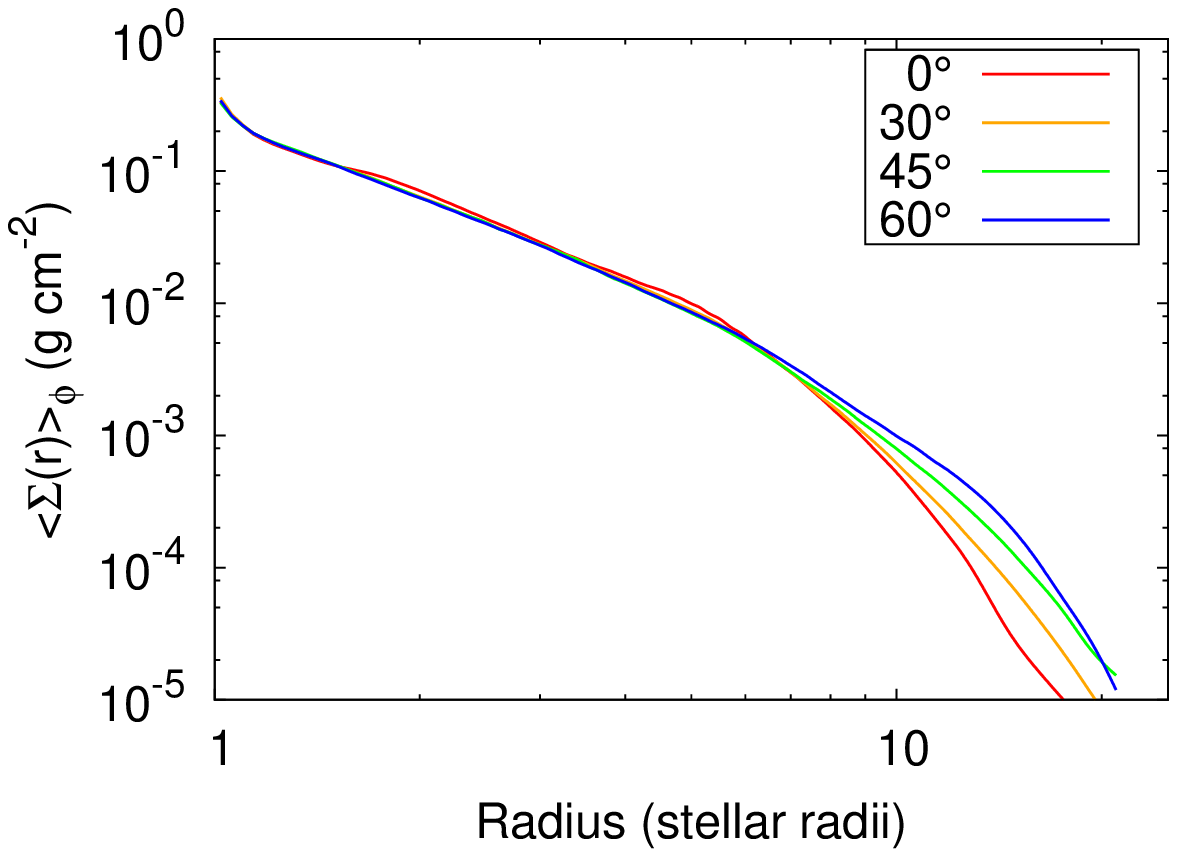}} 
\subfigure[60 day period, $\alpha_\mathrm{SS}=1.0$]{\includegraphics[width = 0.49\columnwidth]{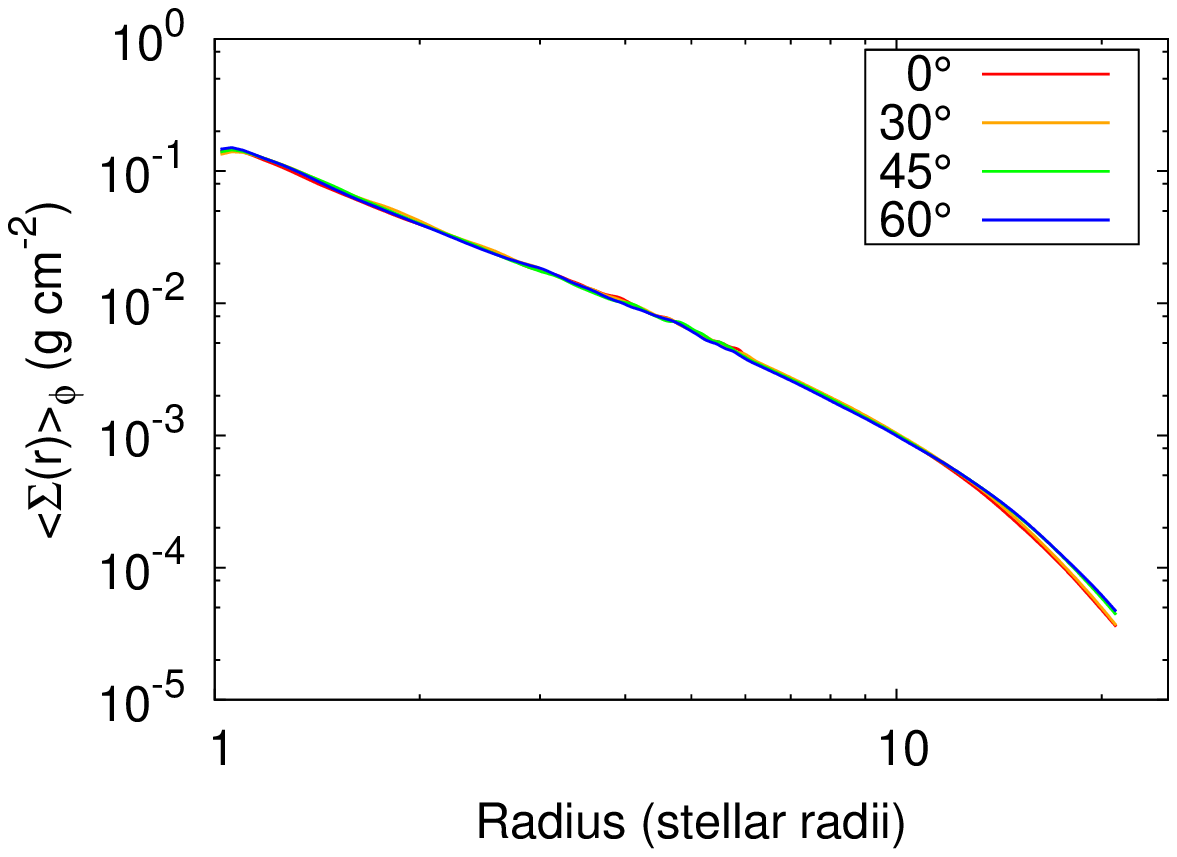}} 
\caption{Azimuthally averaged surface density, $\left < \Sigma(r)\right > _{\phi}$ for the short (left column) and long (right column) orbital period systems, and for $\alpha_\mathrm{SS}$ values of 0.1 (top row), 0.5 (middle row), and 1.0 (bottom row). The results shown were taken 50 $P_{orb}$ after the start of the disc building phase. Each line represents a different misalignment angle, as indicated in the legend.}
\label{fig:avrg_sigma}
\end{figure}

Figures~\ref{fig:avrg_sigma}(a) through (f) show $\left < \Sigma(r)\right > _{\phi}$ after 50 full orbital periods. We see that the $\left < \Sigma(r)\right > _{\phi}$ profiles in each case are similar in shape as those found in \citet{oka02} in his study of Be/X-ray binaries, with shallower slopes in the inner part of the disc and steeper slopes in the outer parts. As \citet{oka02} explains this drop in surface density, or truncation, is due to the tidal forces generated by the secondary as it orbits the primary. Furthermore, similar to the findings of \citet{oka02}, we notice that surface density profiles of the outer part of the disc become shallower with increasing $\alpha_\mathrm{SS}$. This can be attributed to the fact that high viscosity discs have a faster recovery time than low viscosity discs, since viscosity is the primary means with which angular momentum in the disc is transported away from the star.

Now we move onto a systematic investigation of the changes in the disc over the range in parameters described previously. The misalignment angle also affects the surface density profile of the disc as seen in Figure~\ref{fig:avrg_sigma}. In all six panels we see that the greater the misalignment angle, the denser the outermost part of the disc. Conversely, the density in the inner parts appears to be mostly unaffected by the misaligned companion. In order to better investigate this, three properties of these profiles were examined, the truncation radius and the fall-off rate of the surface density in the inner and outer parts of the disc. The inner and outer parts are defined with respect to the truncation radius, the inner disc being the region inside the truncation radius and the outer part exterior to it. For this analysis the azimuthally averaged surface density profiles were fitted using the same function adopted by \citet{oka02}.
\begin{equation}
\left < \Sigma(r)\right > _{\phi} \propto \frac{(r/R_t)^{-m}}{1+(r/R_t)^{n-m}},
\label{eq:fit_rt}
\end{equation}
where $R_t$ is the truncation radius, \textit{m} is the power-law exponent of the inner disc, and \textit{n} is the power-law exponent of the outer disc. Hereafter we will refer to \textit{m} and \textit{n} as the inner and outer exponent, respectively. These three parameters are obtained from this fit.

\begin{figure}
\subfigure[30 day period, $\alpha_\mathrm{SS}=0.1$]{\includegraphics[width = 0.49\columnwidth]{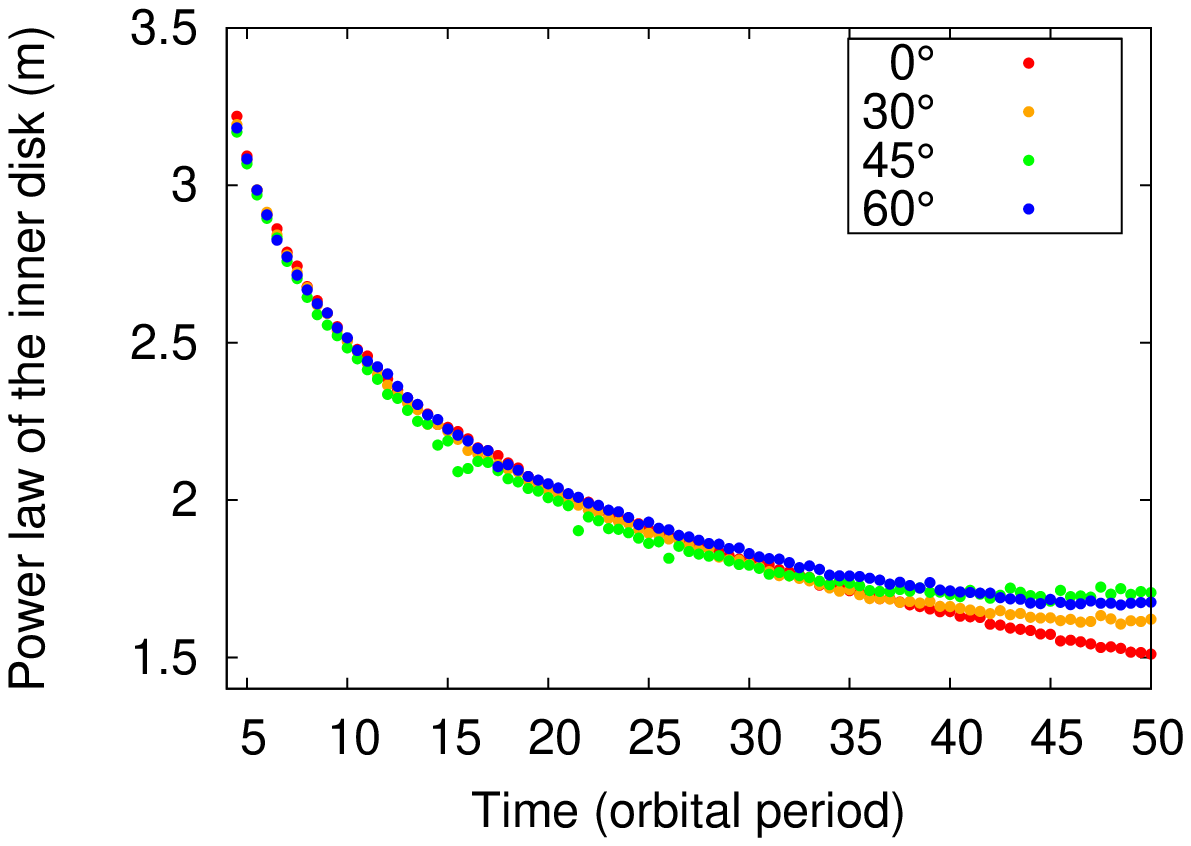}} 
\subfigure[60 day period, $\alpha_\mathrm{SS}=0.1$]{\includegraphics[width = 0.49\columnwidth]{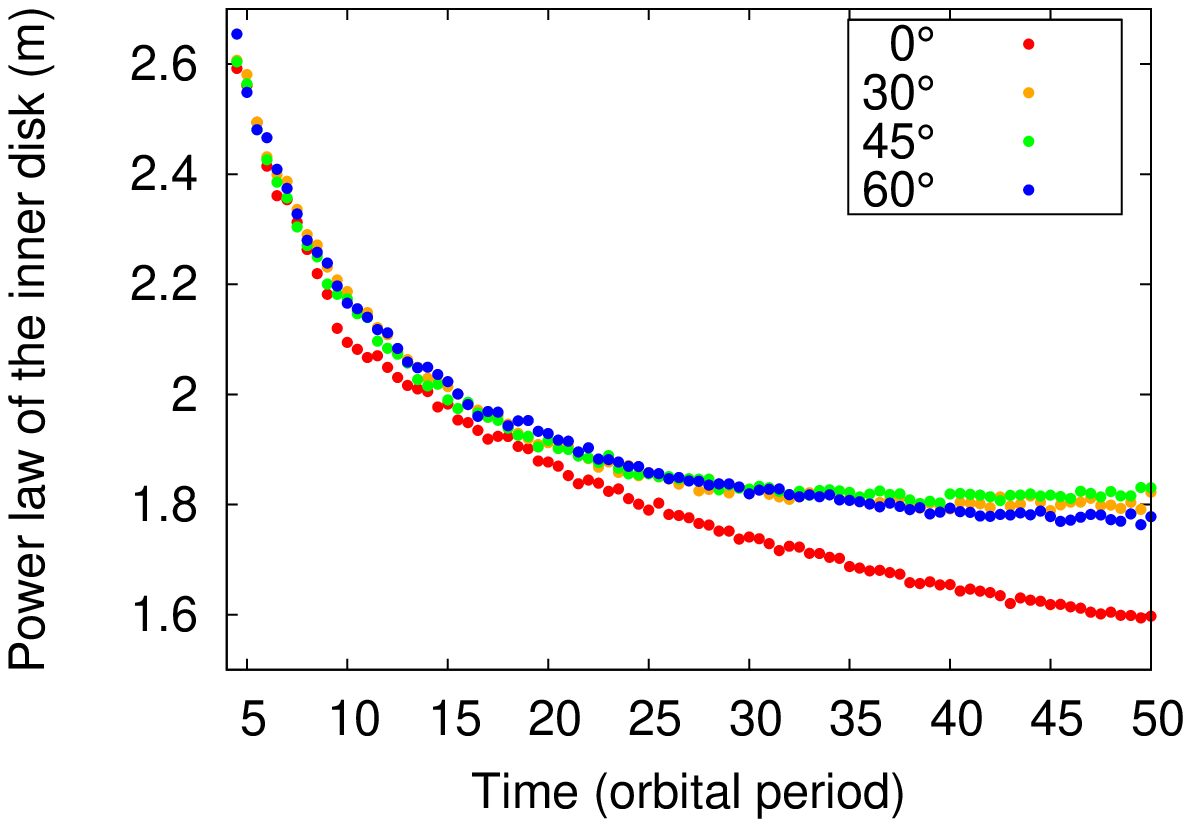}} \\
\subfigure[30 day period, $\alpha_\mathrm{SS}=0.5$]{\includegraphics[width = 0.49\columnwidth]{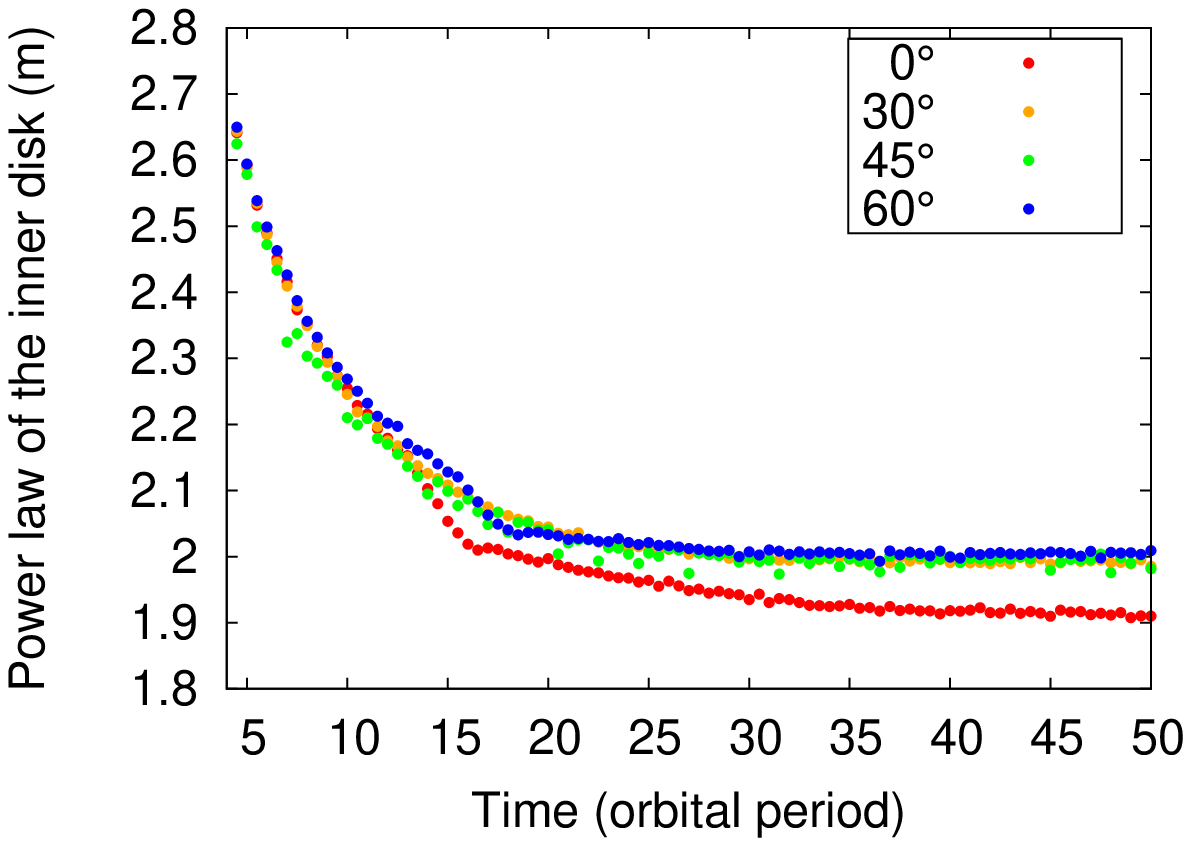}} 
\subfigure[60 day period, $\alpha_\mathrm{SS}=0.5$]{\includegraphics[width = 0.49\columnwidth]{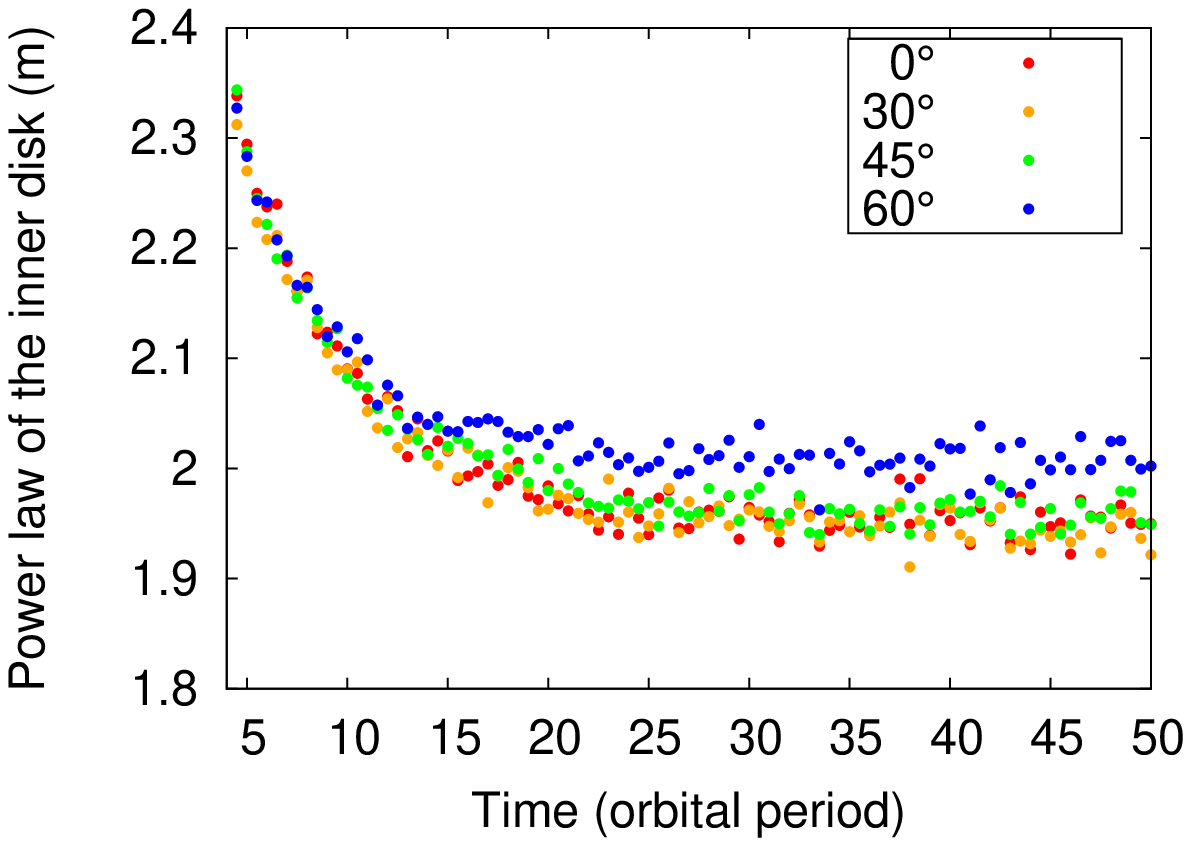}} \\
\subfigure[30 day period, $\alpha_\mathrm{SS}=1.0$]{\includegraphics[width = 0.49\columnwidth]{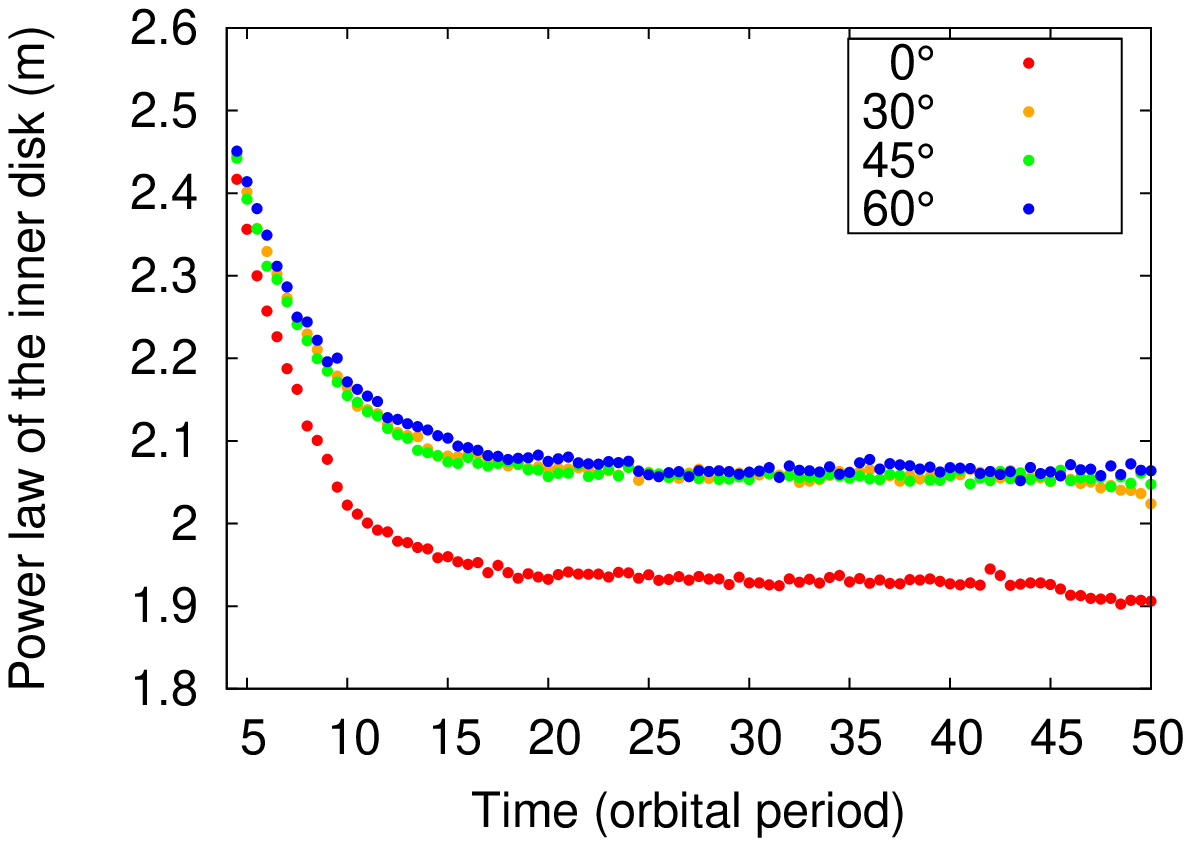}} 
\subfigure[60 day period, $\alpha_\mathrm{SS}=1.0$]{\includegraphics[width = 0.49\columnwidth]{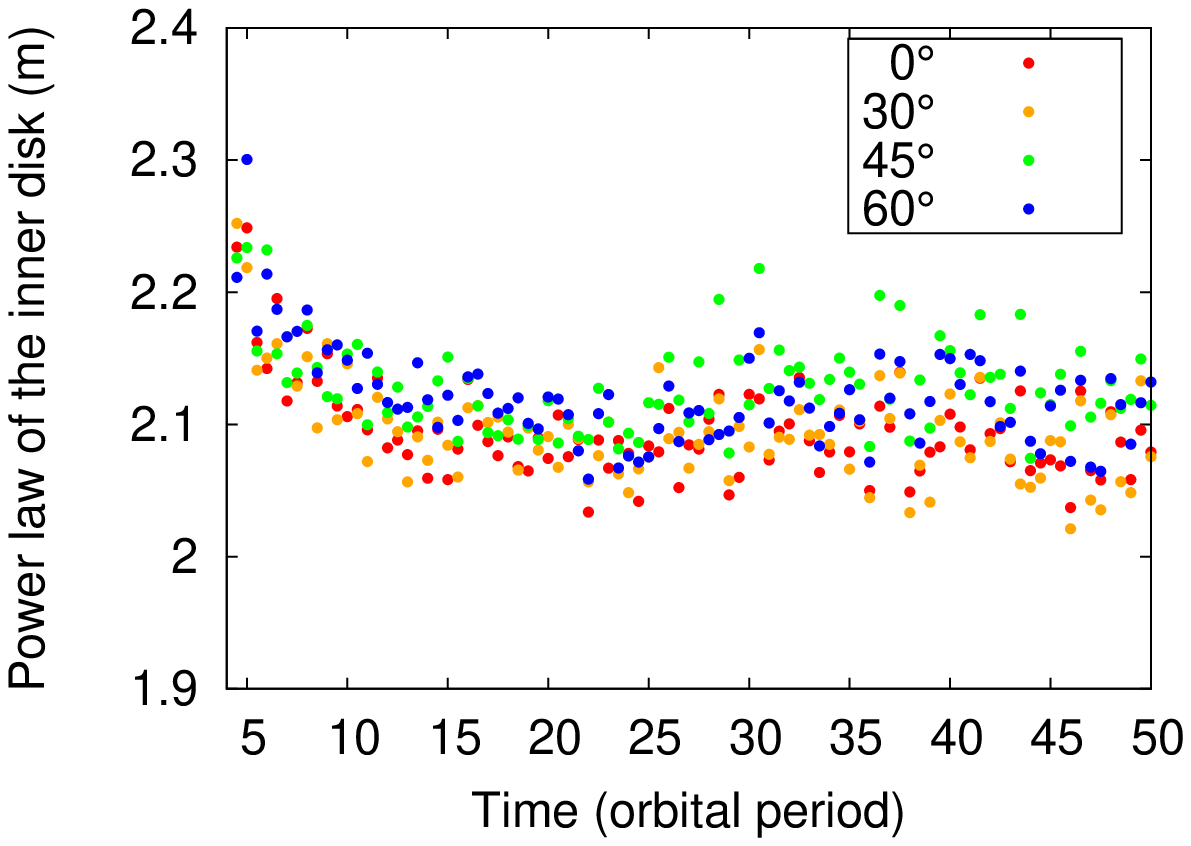}} 
\caption{Temporal evolution of the slope of the inner disc, obtained from Equation~\ref{eq:fit_rt} for the short (left column) and long (right column) orbital period systems, and for $\alpha_\mathrm{SS}$ values of 0.1 (top row), 0.5 (middle row), and 1.0 (bottom row).}
\label{fig:inner}
\end{figure}

Figure~\ref{fig:inner} shows the temporal evolution of the surface density fall-off rate, $m$, in the inner parts of the disc. We see that the fall-off rate is initially high in most cases, but rapidly decreases until a steady-state value is reached. The estimated steady-state values for $m$, obtained by an exponential fit, can be found in Tables~\ref{table:fit_short} and \ref{table:fit_long} for the short and long period systems, respectively.

However, where \citet{hau12} found that the power law for isolated Be-stars tended toward $m=2$, we found, in almost all cases, that the power law tended toward values of $m \leq 2$, which indicates the inner part is denser than a typical disc in an isolated Be star system. The same accumulation effect was also observed in Paper I for aligned systems. As we see from Figure~\ref{fig:inner} as well as Tables~\ref{table:fit_short} and \ref{table:fit_long}, the accumulation effect is stronger, overall, in aligned systems compared to misaligned systems, with little variation in the latter. We notice that the difference in accumulation between the aligned and misaligned systems increases substantially with viscosity.  We also see that both the viscosity coefficient and orbital period have effects on the steady-state value of $m$. These dependences are in agreement with the results of Paper I. The only system where this effect is absent is in the system with both high viscosity and long orbital period (panel f). Here we see that $m$ tends toward values equal or greater than 2. This indicates that the secondary is distant and viscosity is large enough that the torque is too weak to affect the density distribution in the inner disc, which then behaves like an isolated Be disc.

\begin{figure}
\subfigure[30 day period, $\alpha_\mathrm{SS}=0.1$]{\includegraphics[width = 0.49\columnwidth]{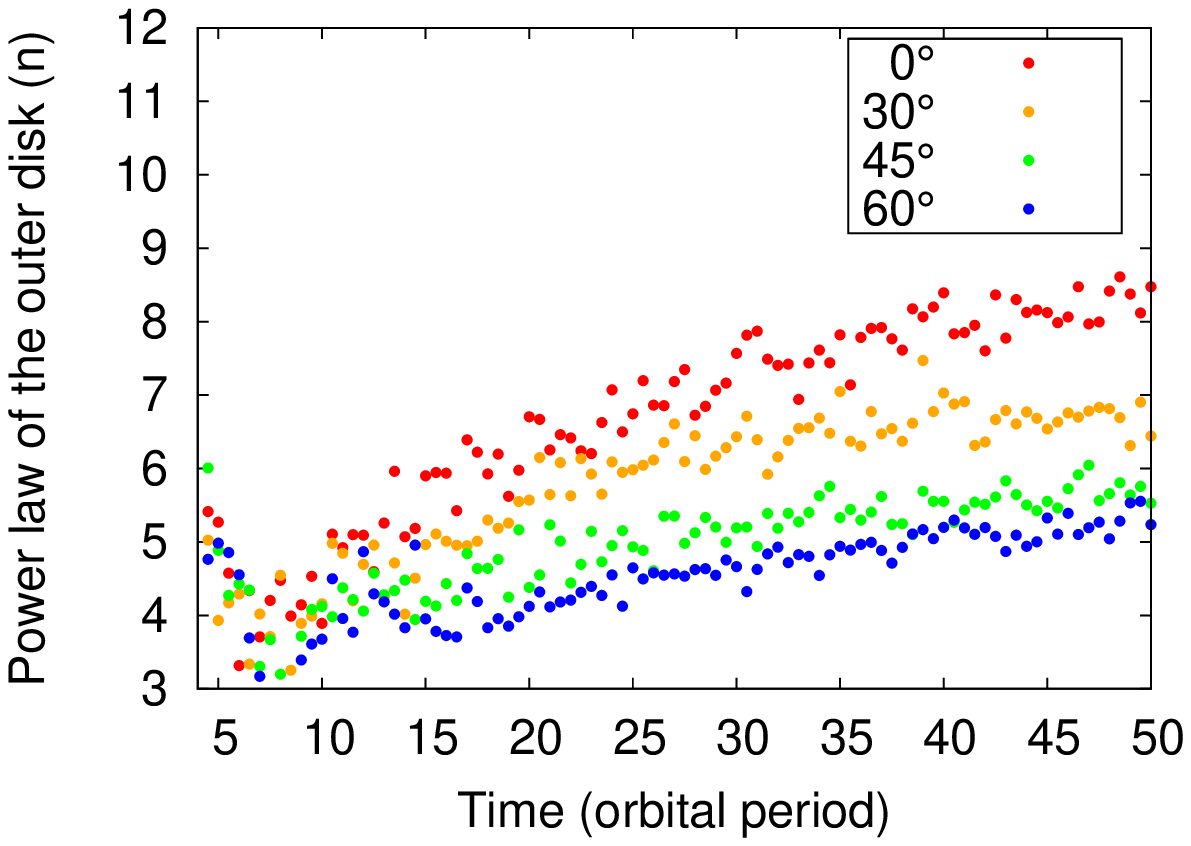}} 
\subfigure[60 day period, $\alpha_\mathrm{SS}=0.1$]{\includegraphics[width = 0.49\columnwidth]{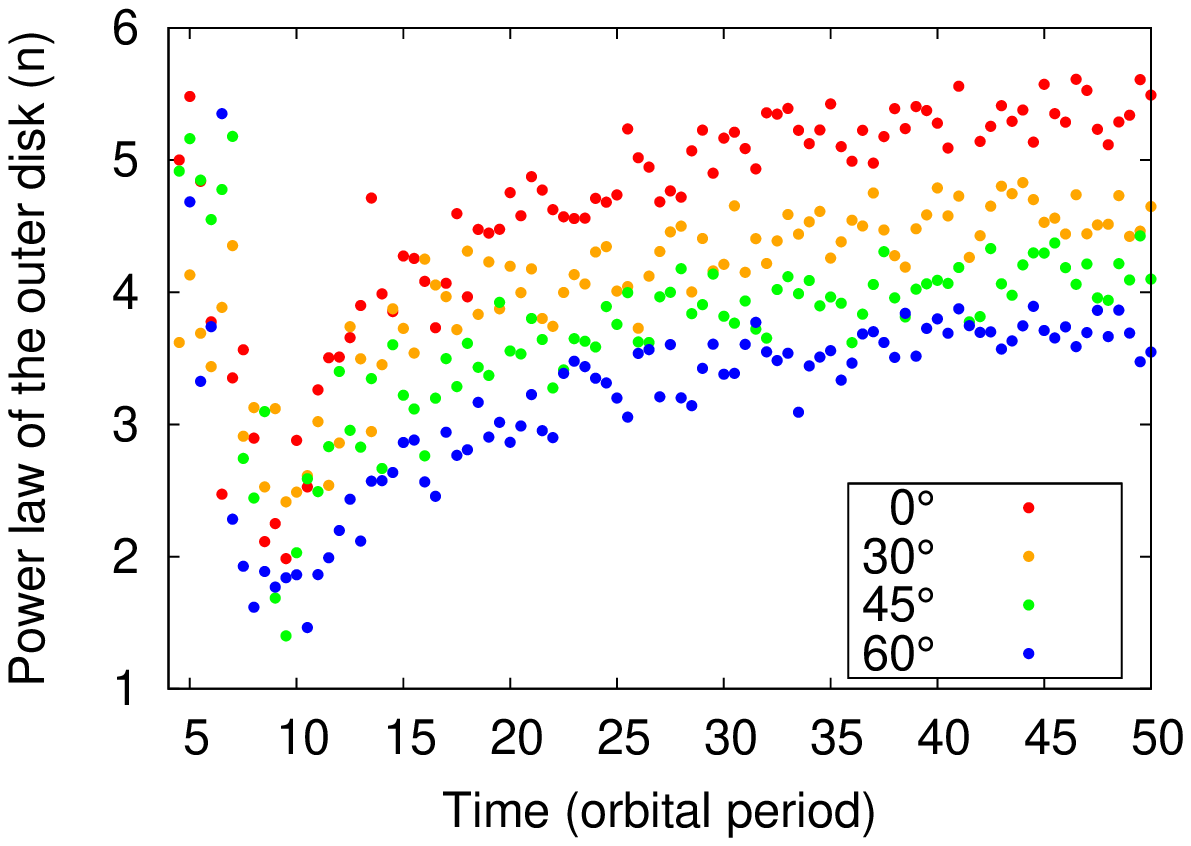}} \\
\subfigure[30 day period, $\alpha_\mathrm{SS}=0.5$]{\includegraphics[width = 0.49\columnwidth]{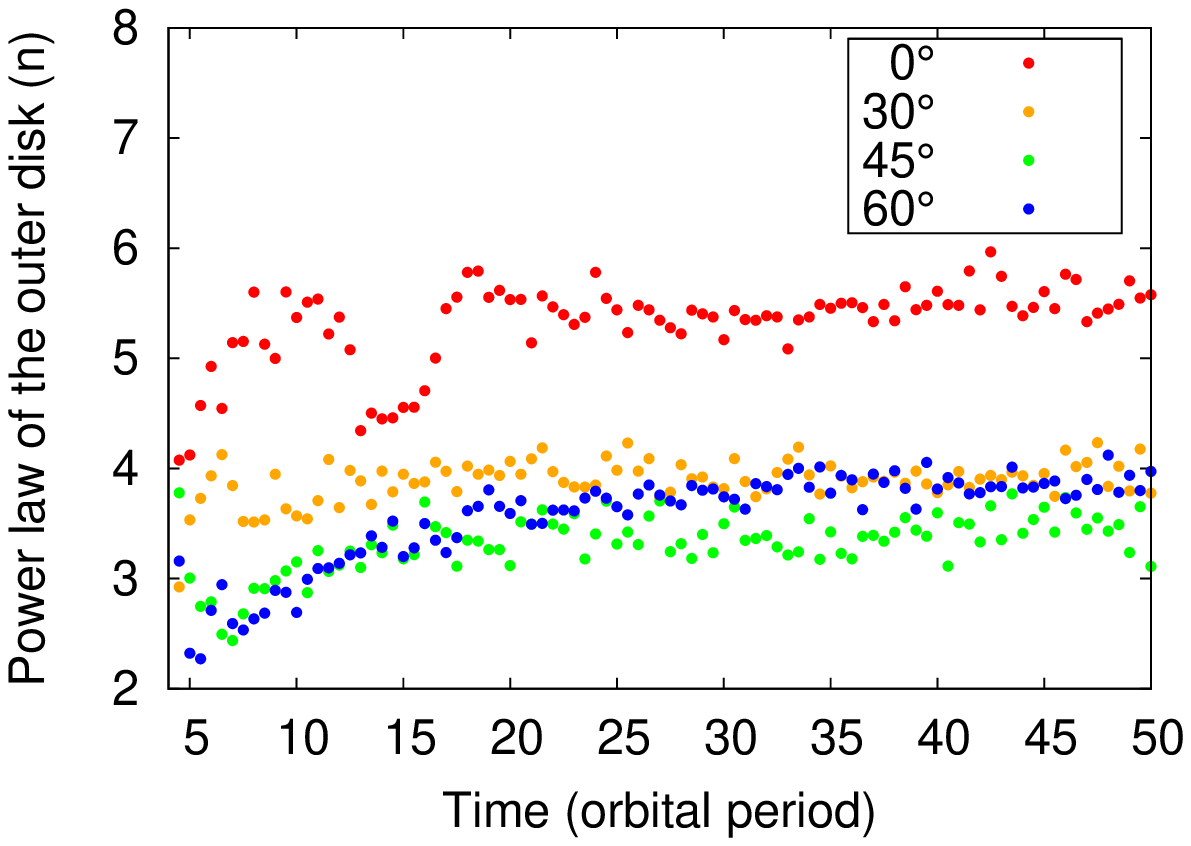}} 
\subfigure[60 day period, $\alpha_\mathrm{SS}=0.5$]{\includegraphics[width = 0.49\columnwidth]{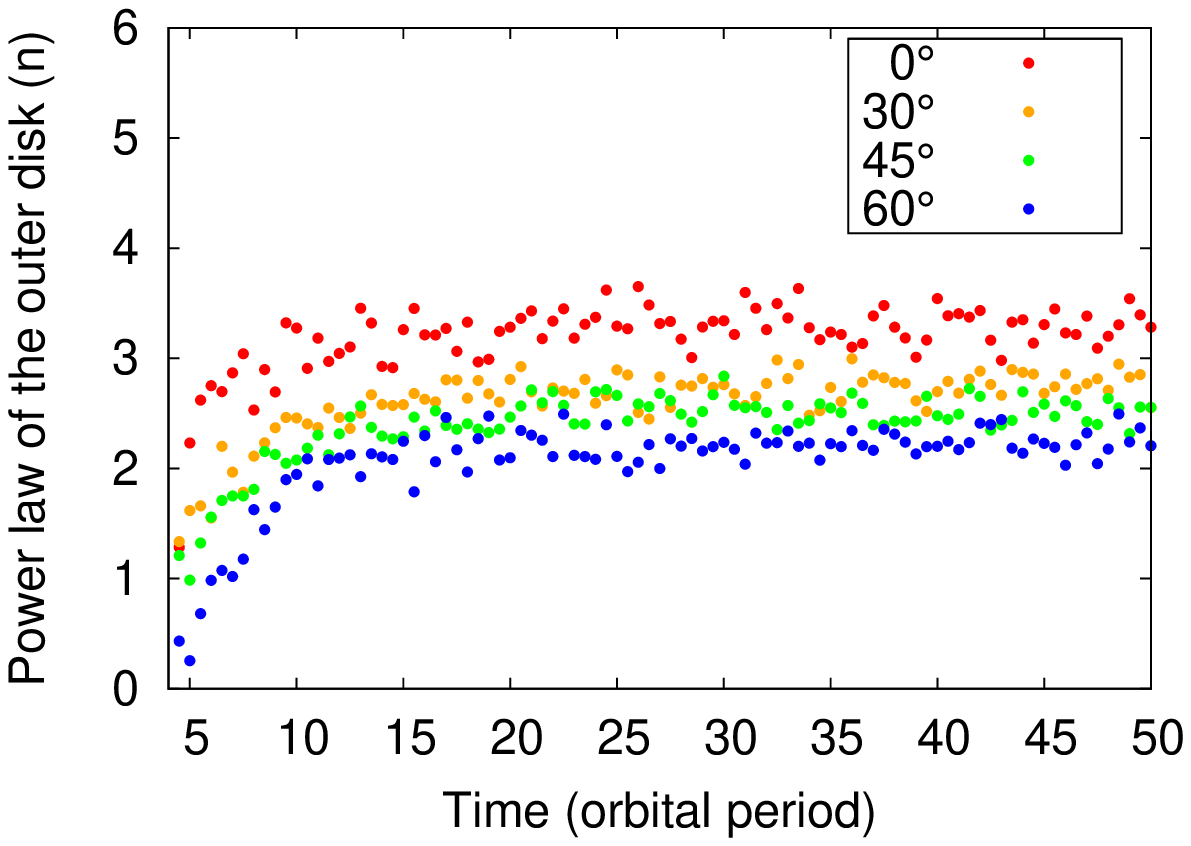}} \\
\subfigure[30 day period, $\alpha_\mathrm{SS}=1.0$]{\includegraphics[width = 0.49\columnwidth]{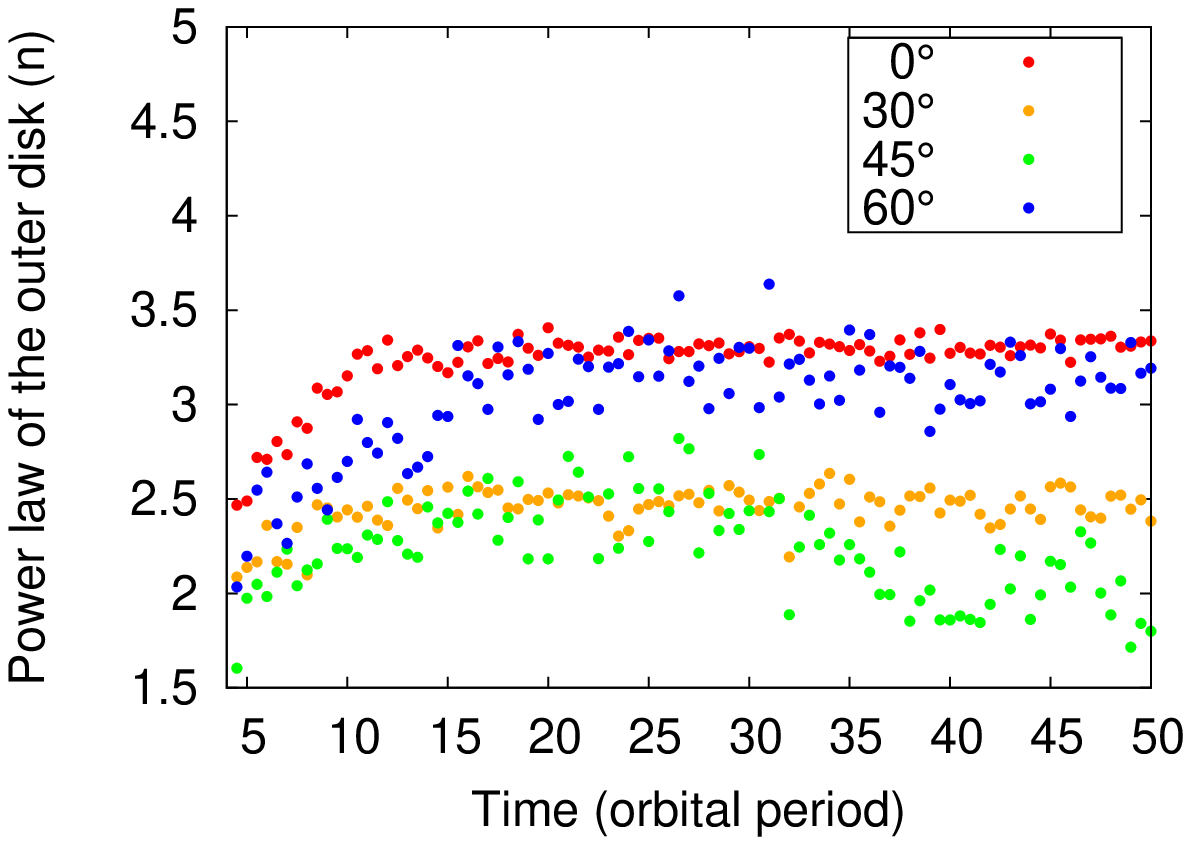}} 
\subfigure[60 day period, $\alpha_\mathrm{SS}=1.0$]{\includegraphics[width = 0.49\columnwidth]{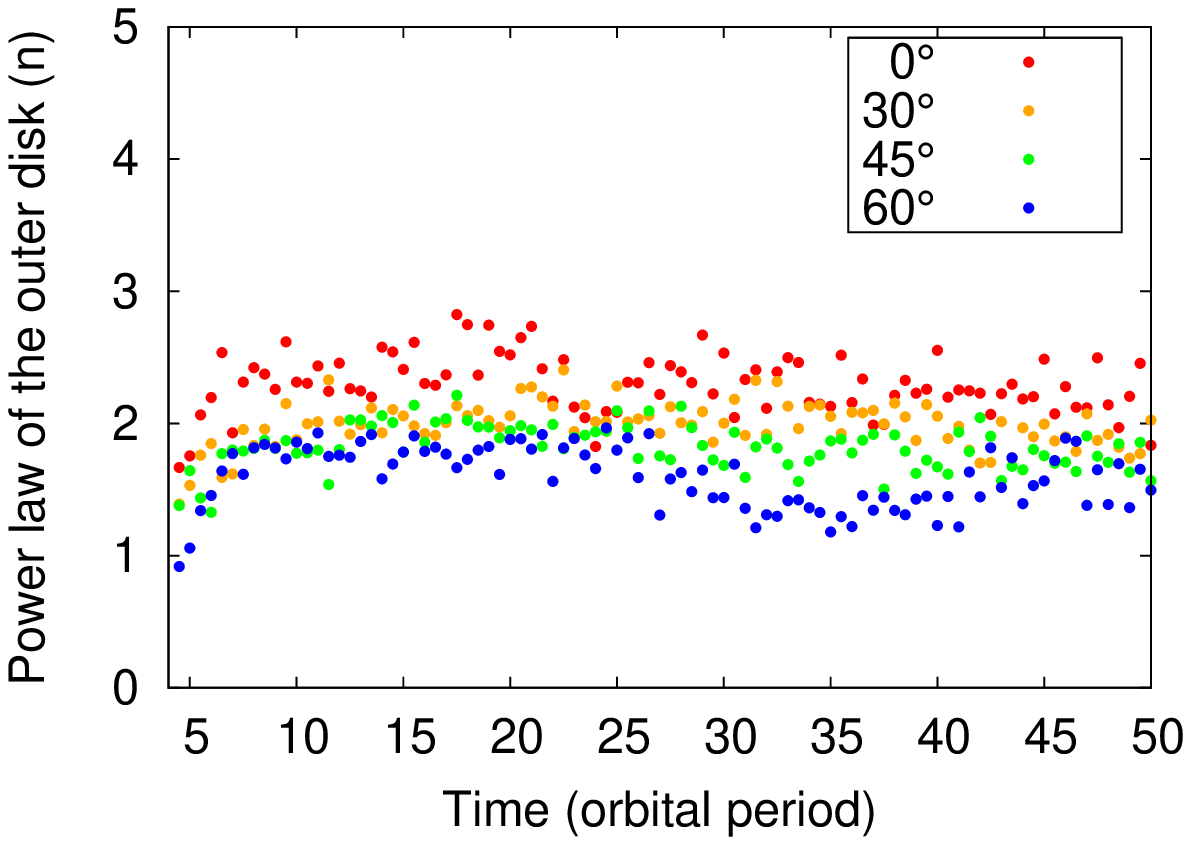}} 
\caption{Same as Figure~\ref{fig:inner} for the slope of the outer disc obtained from Equation~\ref{eq:fit_rt}.}
\label{fig:outer}
\end{figure}

Figure~\ref{fig:outer} shows the temporal evolution of the surface density fall-off rate, $n$, in the outer parts of the disc. We see here that the slope $n$ reaches a steady state value more quickly than $m$. However it is important to note that although the outer density slope reaches QSS faster, the density scale still requires a longer time to reach QSS compared to the inner disc. The estimated steady-state values for $n$ are also presented in Tables~\ref{table:fit_short} and \ref{table:fit_long}. As seen in Figures~\ref{fig:avrg_sigma} and \ref{fig:outer} the fall-off rate in the outer disc is dependent on the misalignment angle, i.e. slower fall-off rates for highly misaligned systems. 

\begin{figure}
\subfigure[30 day period, $\alpha_\mathrm{SS}=0.1$]{\includegraphics[width = 0.49\columnwidth]{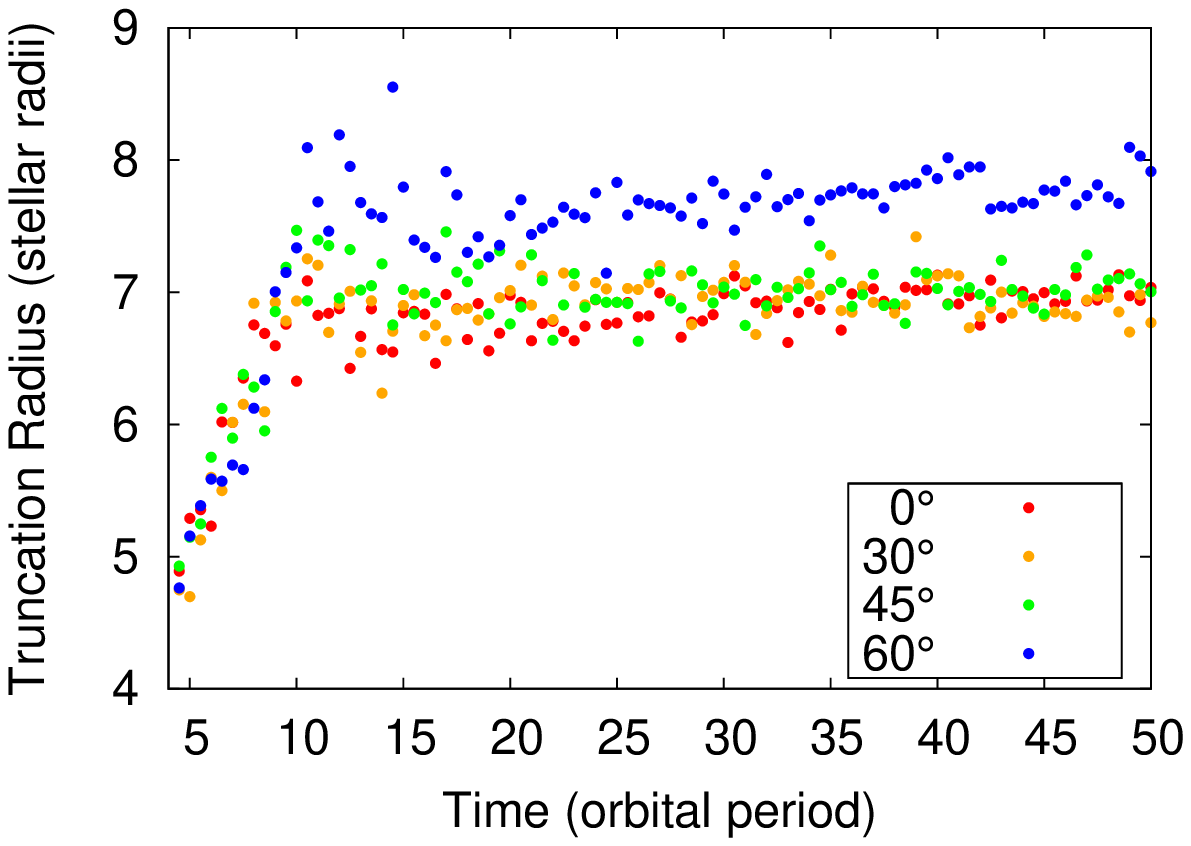}} 
\subfigure[60 day period, $\alpha_\mathrm{SS}=0.1$]{\includegraphics[width = 0.49\columnwidth]{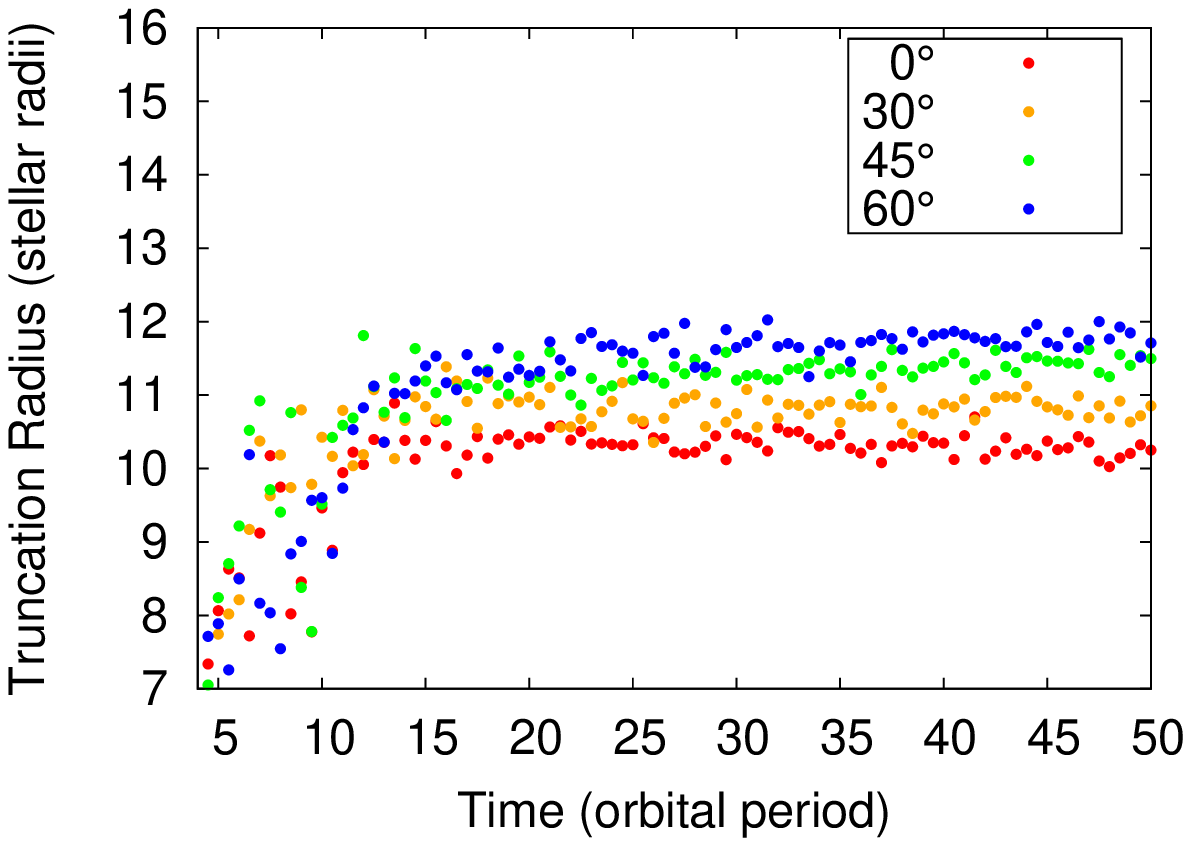}} \\
\subfigure[30 day period, $\alpha_\mathrm{SS}=0.5$]{\includegraphics[width = 0.49\columnwidth]{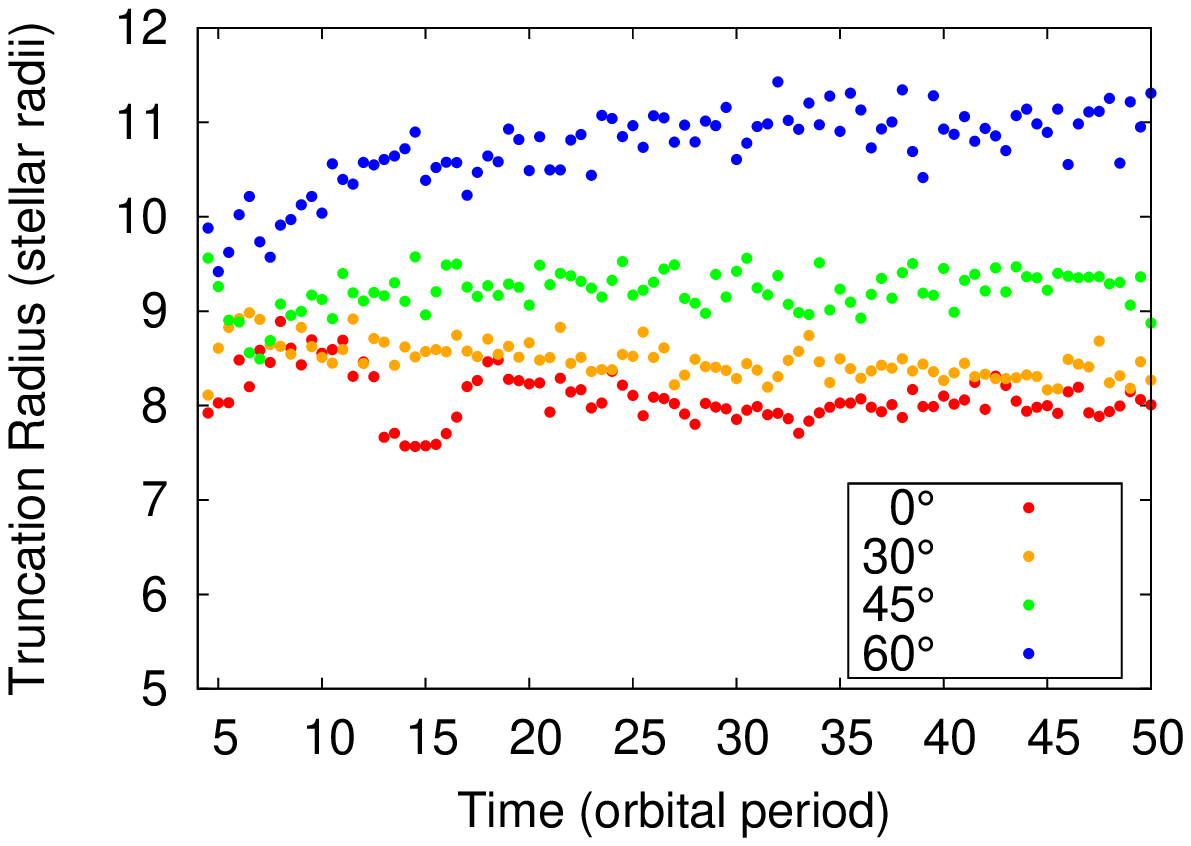}} 
\subfigure[60 day period, $\alpha_\mathrm{SS}=0.5$]{\includegraphics[width = 0.49\columnwidth]{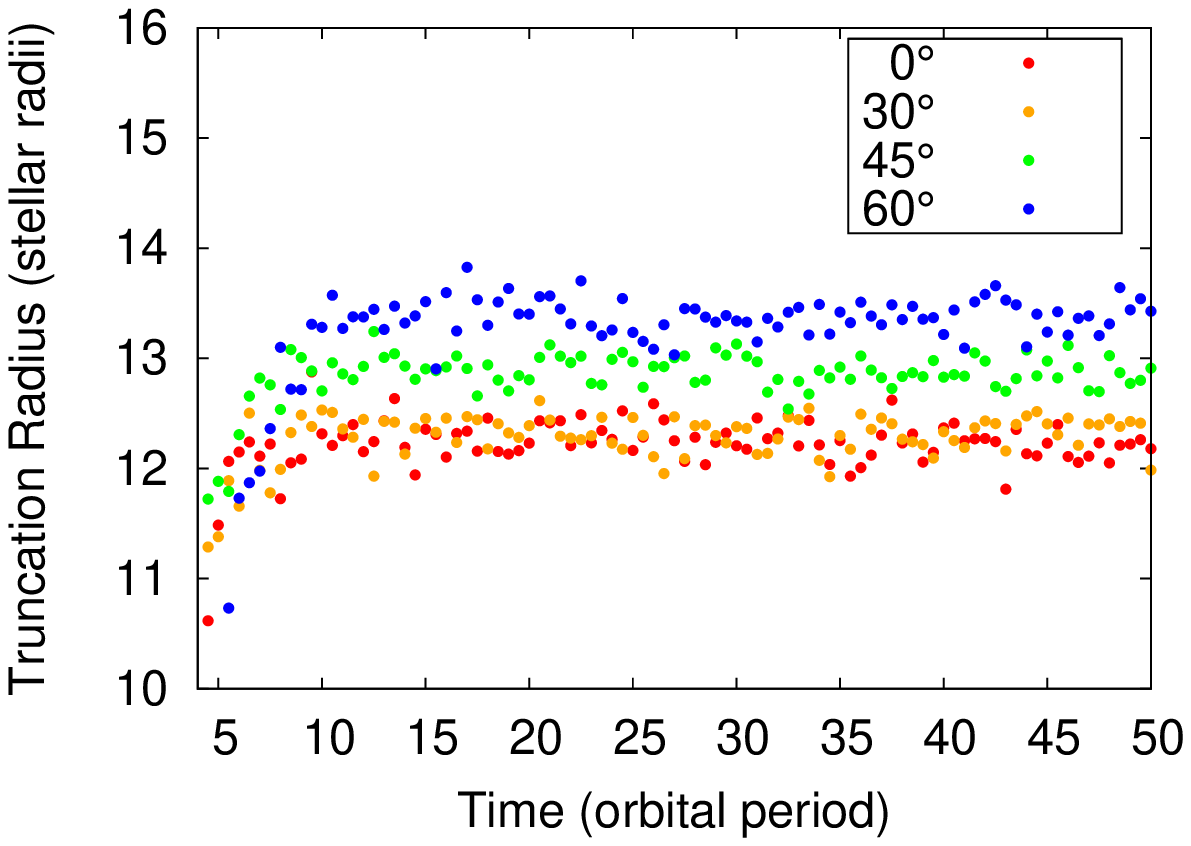}} \\
\subfigure[30 day period, $\alpha_\mathrm{SS}=1.0$]{\includegraphics[width = 0.49\columnwidth]{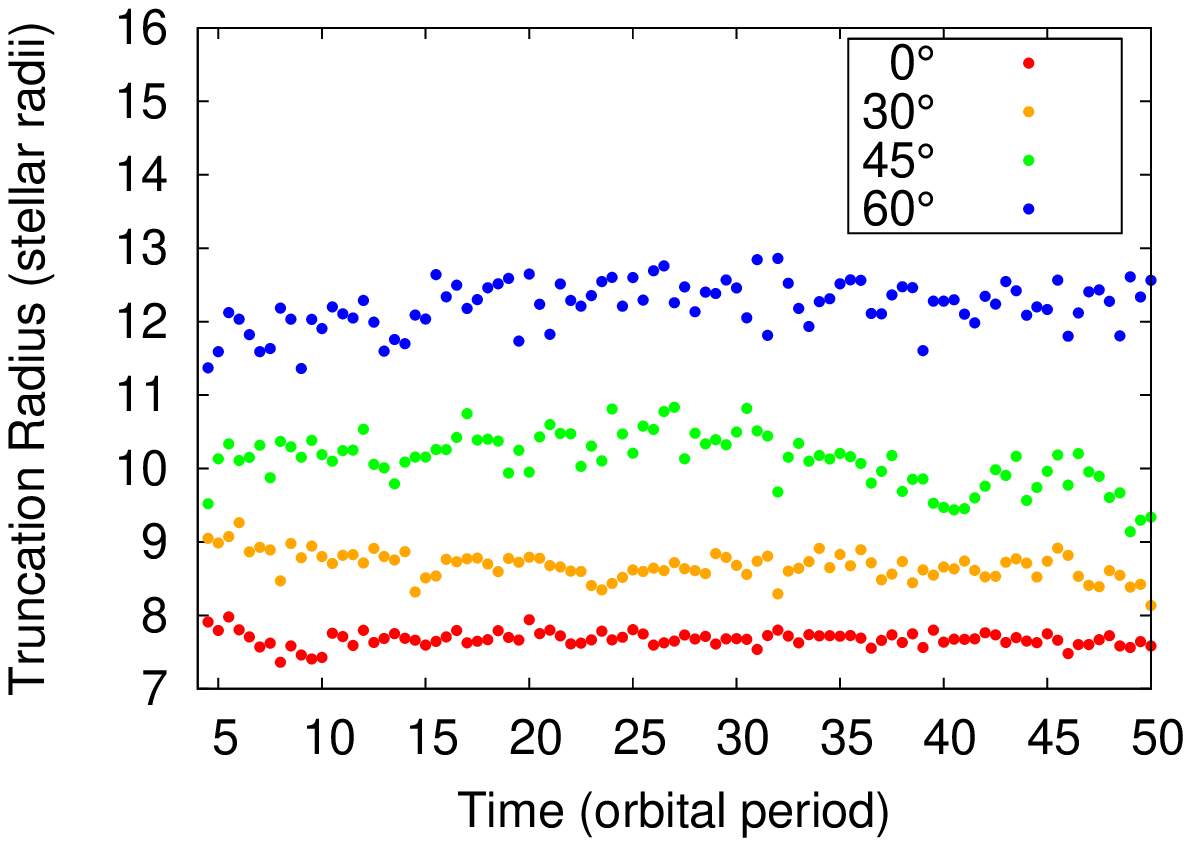}} 
\subfigure[60 day period, $\alpha_\mathrm{SS}=1.0$]{\includegraphics[width = 0.49\columnwidth]{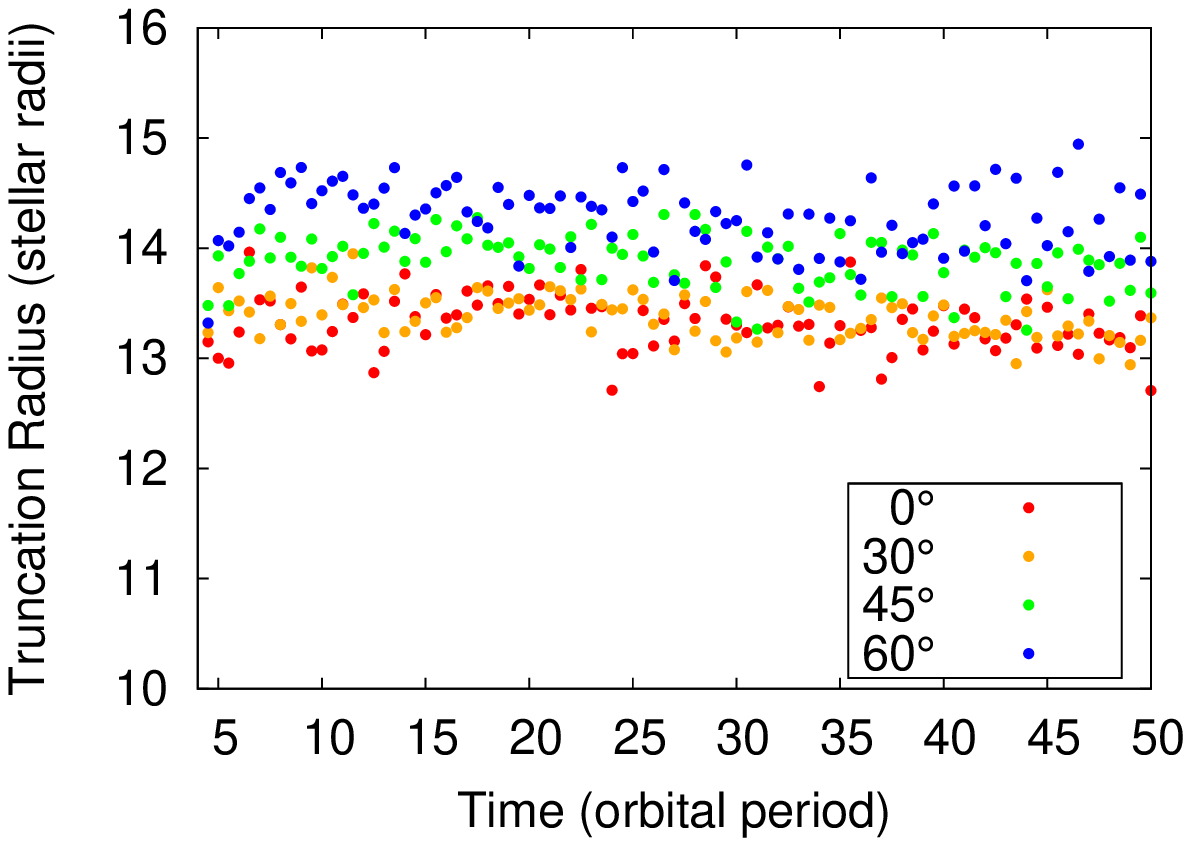}} 
\caption{Same as Figure~\ref{fig:inner} for the truncation radius obtained from Equation~\ref{eq:fit_rt}.}
\label{fig:rt}
\end{figure}

The results for the truncation radius, obtained from Equation~\ref{eq:fit_rt}, are shown in Figure~\ref{fig:rt} with the steady-state values in Tables~\ref{table:fit_short} and \ref{table:fit_long}. The panel layout is the same as Figures~\ref{fig:avrg_sigma} and \ref{fig:outer}. We see that the steady-state values of the truncation radius tends to be closer to the star for low misalignment angles and move farther out as the misalignment angle increases. We also notice that the truncation radius is farther away from the star in systems with higher disc viscosity and longer orbital periods. This is similar to the findings presented in Paper I for aligned systems. 

The relationship between $R_t$, $m$, and $n$ with regards to viscosity, orbital period, and misalignment angle is relatively intuitive. As mentioned earlier, viscosity determines how fast the disc can grow and recover between each interaction with the secondary. We therefore expect high viscosity discs to have more time to relax back to their unperturbed state than less viscous ones, reducing the impact of the accumulation effect (higher $m$) and allowing more material to flow into the outer disc (lower $n$ and higher $R_t$). A longer orbital period also means that the outer disc has more time the rebuild itself between passages, and since the secondary is radially farther away from the disc, it has a smaller impact on the truncation of the disc. Similarly, increasing the misalignment angle also increases the vertical distance between the disc and the secondary when the latter is at its maximum elevation. This increased separation weakens the tidal torque on the disc resulting in a weaker truncation.

\begin{table*}
 \caption{Steady-state values for the short period (30 day) systems.}
 \label{table:fit_short}
 \begin{tabular}{ccccc}
  \hline
  Viscosity & Misalignment  & Inner slope  & Outer slope  & Truncation radius\\
  parameter & angle  & $m$  & $n$  & $R_T$ (stellar radii)\\
  \hline
   & 0$\degr$ & 1.49 & 7.78 & 6.91\\
$\alpha_\mathrm{SS} = 0.1$  & 30$\degr$ & 1.58 & 6.60 & 6.97\\
 & 45$\degr$ & 1.67 & 5.42 & 7.00\\
 & 60$\degr$ & 1.65 & 4.97 & 7.72\\
  \hline
 & 0$\degr$ & 1.92 & 5.48 & 8.03\\
$\alpha_\mathrm{SS} = 0.5$  & 30$\degr$ & 1.99 & 3.95 & 8.41\\
 & 45$\degr$ & 1.99 & 3.43 & 9.27\\
 & 60$\degr$ & 2 & 3.81 & 10.9\\
  \hline
   & 0$\degr$ & 1.93 & 3.31 & 7.68\\
$\alpha_\mathrm{SS} = 1.0$ & 30$\degr$ & 2.06 & 2.47 & 8.62\\
 & 45$\degr$ & 2.06 & 2.22 & 10.1\\
 & 60$\degr$ & 2.06 & 3.17 & 12.3\\

\hline
 \end{tabular}
\end{table*}

\begin{table*}
 \caption{Steady-state values for the long period (60 day) systems.}
 \label{table:fit_long}
 \begin{tabular}{ccccc}
  \hline
  Viscosity & Misalignment  & Inner slope  & Outer slope  & Truncation radius\\
  parameter & angle  & $m$  & $n$  & $R_T$ (stellar radii)\\
  \hline
    & 0$\degr$ & 1.61 & 5.23 & 10.34\\
$\alpha_\mathrm{SS} = 0.1$  & 30$\degr$ & 1.8 & 4.41 & 10.81\\
 & 45$\degr$ & 1.81 & 3.96 & 11.34\\
 & 60$\degr$ & 1.78 & 3.55 & 10.69\\
  \hline
 & 0$\degr$ & 1.95 & 3.32 & 12.25\\
$\alpha_\mathrm{SS} = 0.5$  & 30$\degr$ & 1.95 & 2.75 & 12.32\\
 & 45$\degr$ & 1.96 & 2.54 & 12.89\\
 & 60$\degr$ & 2.01 & 0.22 & 13.37\\
  \hline
   & 0$\degr$ & 2.08 & 2.27 & 13.30\\
$\alpha_\mathrm{SS} = 1.0$ & 30$\degr$ & 2.09 & 2.01 & 13.34\\
 & 45$\degr$ & 2.12 & 1.81 & 13.84\\
 & 60$\degr$ & 2.11 & 1.55 & 14.23\\
\hline
 \end{tabular}
\end{table*}

\subsection{Disc Warping}

In order to study the warping of the disc, we consider the vertical displacement of the disc above and below the equatorial plane along various radial directions. To accomplish this, we first determine the position of the centre of mass along the vertical axis ($z_\mathrm{CoM}$) along the $r$-axis for various values of $\phi$ (dashed lines in Figure~\ref{fig:orbit}), using the following equation:

\begin{equation}
z_\mathrm{CoM}(r,\phi)=\frac{\sum_{k}{\rho_k(r,\phi) z_k}}{\sum_{k}{\rho_k(r,\phi)}},
\label{eq:CoM}
\end{equation}
where $\rho_k(r,\phi)$ and $z_k$ are the density and corresponding vertical position of a column density located at $(r,\phi)$. Next we convert the vertical position $z_\mathrm{CoM}$ into an angular position $\gamma$ by trigonometry:

\begin{equation}
\gamma(r,\phi)=\arctan{\frac{z_\mathrm{CoM}(r,\phi)}{r}}.
\label{eq:warp_deg}
\end{equation}

We then calculated the tilt of each ring section of the disc by fitting the following equation using the results obtained by Equation~\ref{eq:warp_deg} for multiple values of $r$:
\begin{equation}
\gamma(\phi)_{r}=\tau_r \sin(\phi + \beta_r),
\label{eq:tilt}
\end{equation}
where $\tau_r$ the tilting angle of the disc at radius $r$, defined as the angle formed by the normal of the disc section and the $z$-axis. Similarly, $\beta$ represents the direction of the line of node of the ring section in relation to the equator of the primary.

\begin{figure}
\includegraphics[width = 0.9\columnwidth]{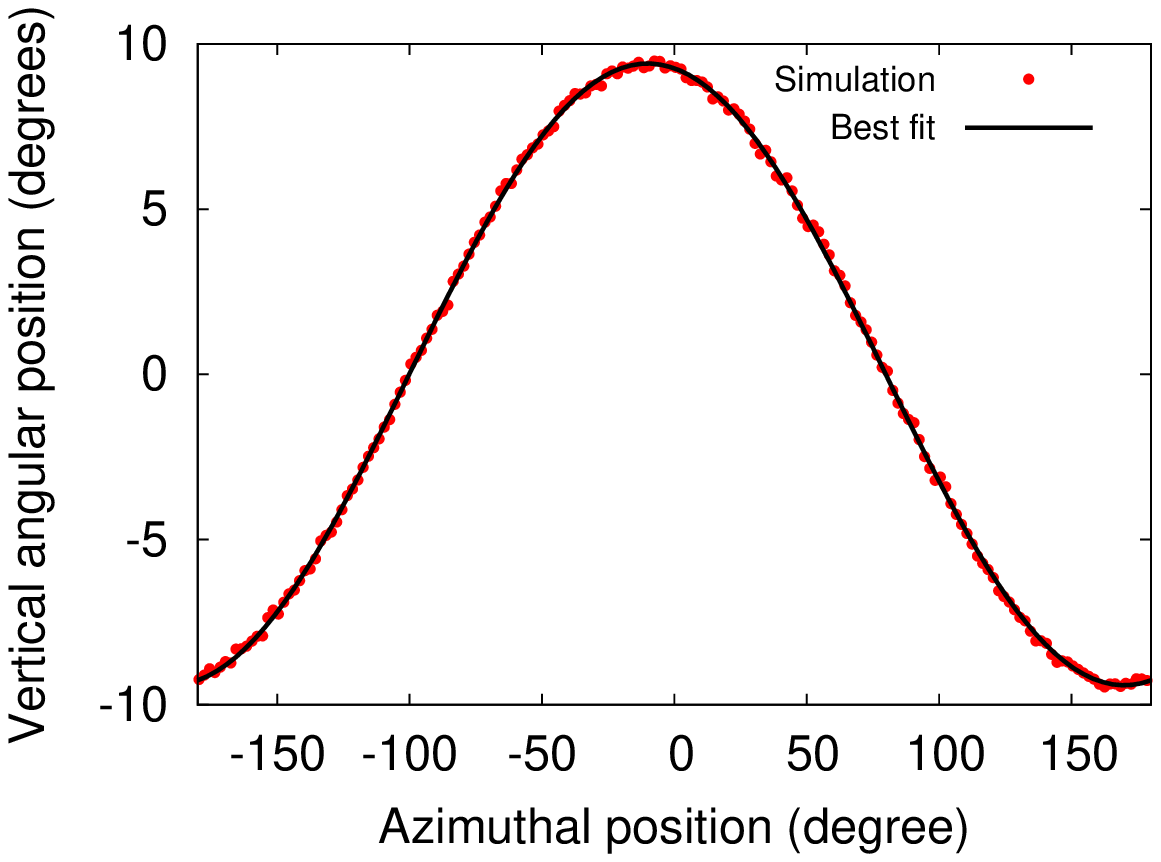} 
\caption{Angular position of the center of mass from the equatorial plane as a function of azimuthal angle at a radial distance of $r=2.79$ stellar radii for the short period system with $\alpha_\mathrm{SS} = 0.5$ and a misalignment angle of 30$\degr$. The red dots represent data taken from the simulation while the solid gray line shows the result of the best fit using Equation~\ref{eq:tilt}.}  
\label{fig:cos_fit}
\end{figure}

An example of this procedure can be found in Figure~\ref{fig:cos_fit}, which shows the relation between the angular position, $\gamma$, of the center of mass, defined from the equatorial plane, and the azimuthal angle, $\phi$, at a radial distance of $r=2.79$ stellar radii. Data taken from the simulation are represented by the black dots while the solid gray line shows the result of the best fit using Equation~\ref{eq:tilt}. Note that, close to the star, the scatter on the data from the simulation was higher due to the constant injection of new mass, however we managed to obtain a good fit.

\begin{figure}
\subfigure[Misalignment angle = 30$\degr$]{\includegraphics[width = 0.9\columnwidth]{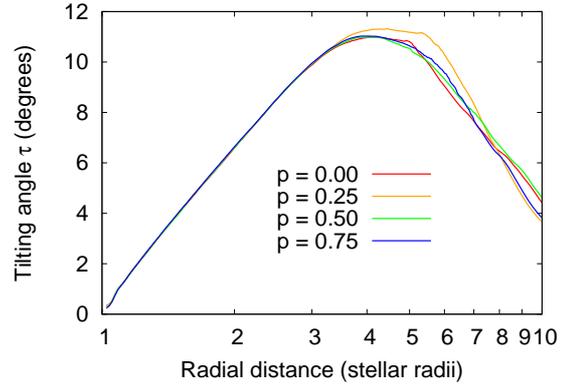}}\\
\subfigure[Misalignment angle = 45$\degr$]{\includegraphics[width = 0.9\columnwidth]{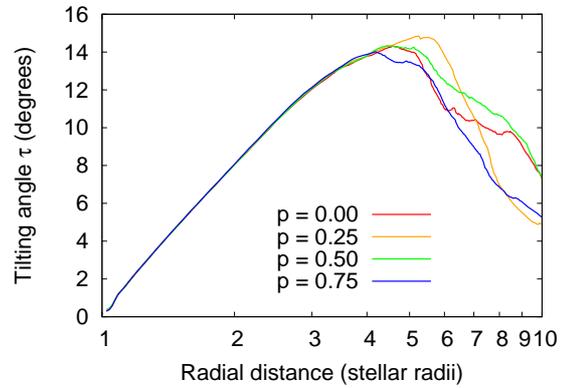}}\\
\subfigure[Misalignment angle = 60$\degr$]{\includegraphics[width = 0.9\columnwidth]{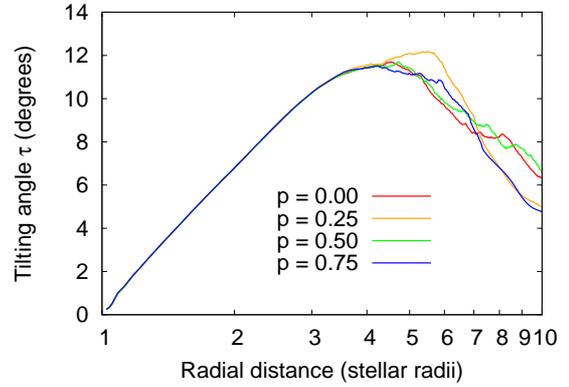}}\\

\caption{Tilting angle of the disc as a function of radius in the short period system with $\alpha_\mathrm{SS} = 0.5$ and a misalignment angle of 30$\degr$ (top), 45$\degr$ (middle), and 60$\degr$ (bottom). The coloured lines represent the orbital phase of the binary, $p$, over one orbital period from 49 $P_{orb}$ to 50 $P_{orb}$; $p=0.0$ (red), $p=0.25$ (orange), $p=0.5$ (green), and $p=0.75$ (blue).}  
\label{fig:tilt}
\end{figure}

Figure~\ref{fig:tilt} shows the tilting angle, $\tau$, as a function of radius at various orbital phases ($p$ = 0.0, 0.25, 0.5, 0.75) for our three misaligned systems.

As expected, we see clear evidence of vertical displacement in the disc as well as other interesting features. In all three cases, the tilting increases steadily in the first few stellar radii until reaching a peak around 4-5 stellar radii. Furthermore this increase does not appear to vary with orbital phase suggesting that the tilting in the very inner region is quite stable. We also notice that the tilt becomes more significant as the misalignment angle increases from 30$\degr$ to 45$\degr$, but then decreases slightly when increased to 60$\degr$.

\begin{figure}
\subfigure[Misalignment angle = 30$\degr$]{\includegraphics[width = 0.9\columnwidth]{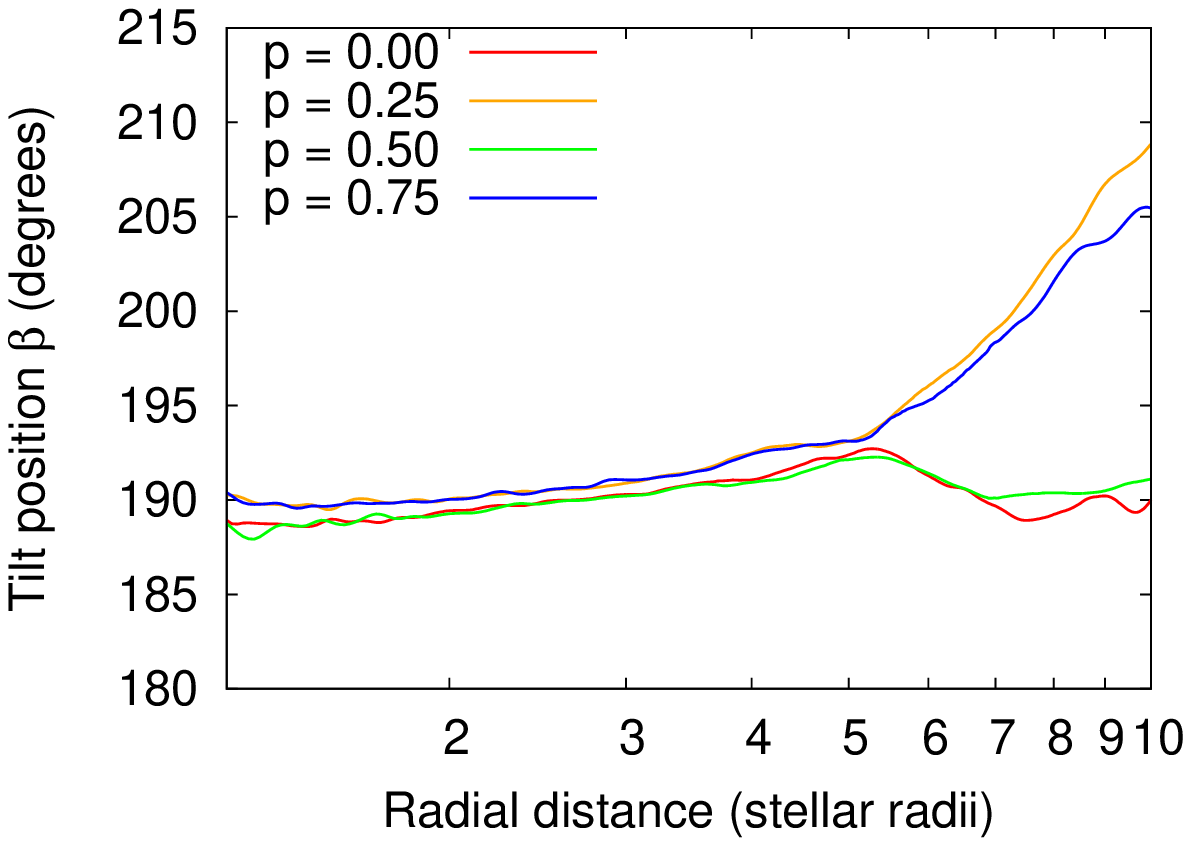}}\\
\subfigure[Misalignment angle = 45$\degr$]{\includegraphics[width = 0.9\columnwidth]{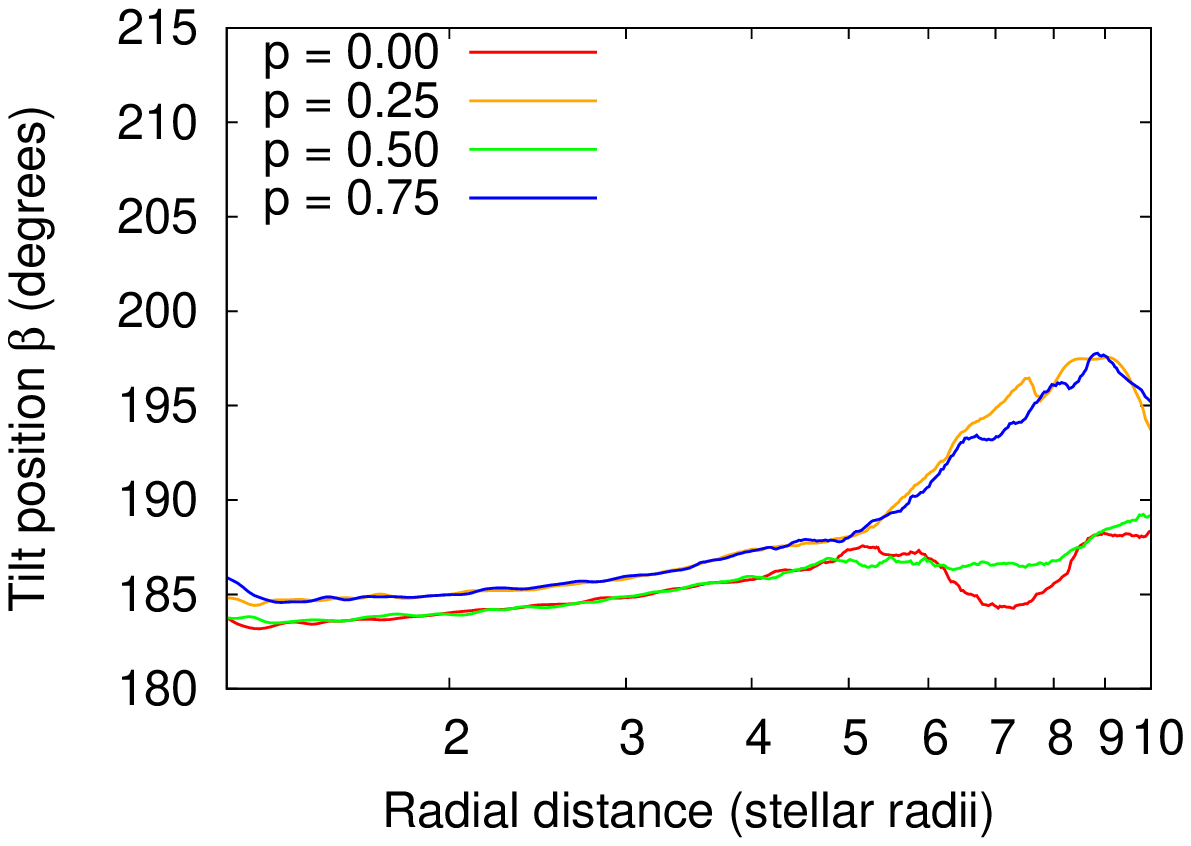}}\\
\subfigure[Misalignment angle = 60$\degr$]{\includegraphics[width = 0.9\columnwidth]{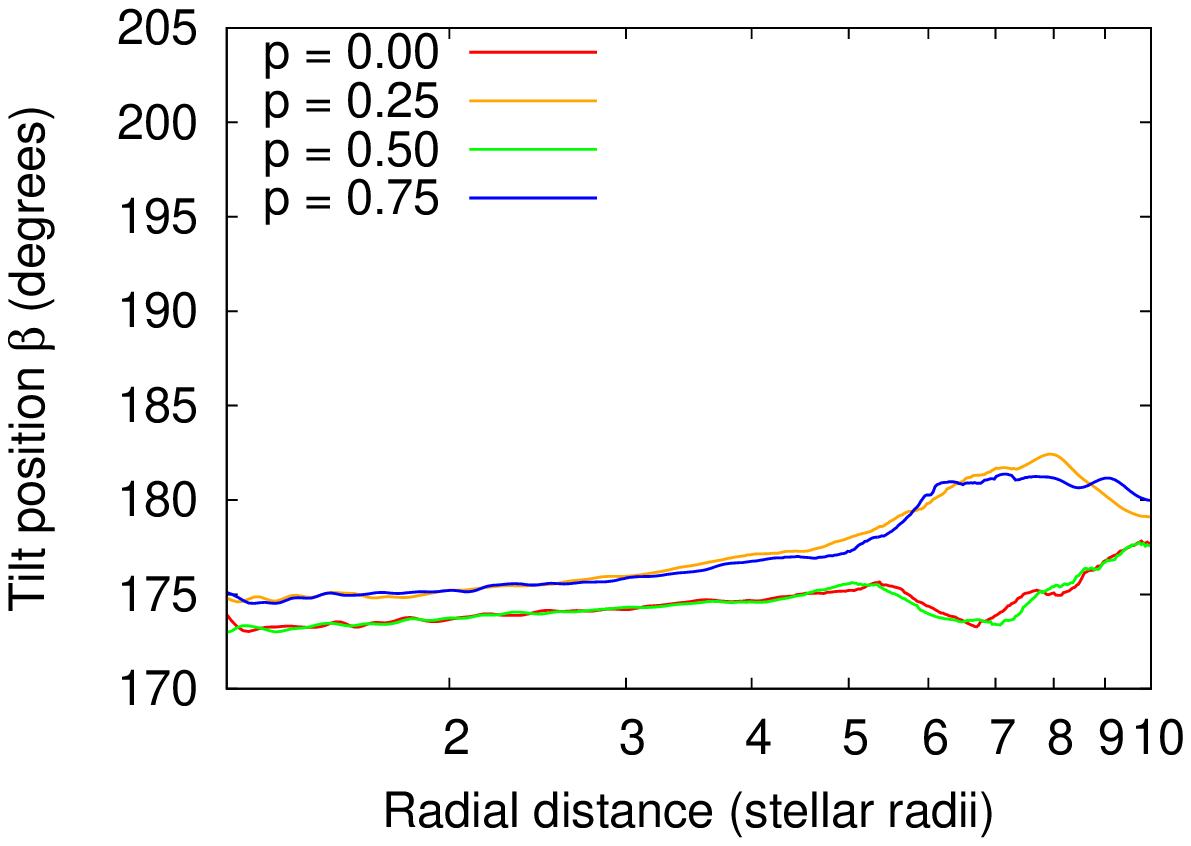}}\\

\caption{Same as Figure~\ref{fig:tilt} except for the direction of the line of nodes.} 
\label{fig:prec}
\end{figure}

Similarly to Figure~\ref{fig:tilt}, Figure~\ref{fig:prec} shows the direction of the line of nodes, $\beta$, as a function of radius at various orbital phases ($p$ = 0.0, 0.25, 0.5, 0.75) for our three misaligned systems. The line of node is defined with respect to the plane of the equator of the central star and its direction is defined in the same way as the azimuthal angle $\phi$, as depicted in Figure~\ref{fig:orbit}. Also note that due to we will only use one angular value to define lines of nodes to avoid confusion. We notice that the disc sections tend to tilt toward $\phi =$ 80$\degr$ to 100$\degr$. This is an offset of about 90$\degr$ compared to the line of nodes of the orbital plane, which is $\phi = 0\degr$  for all misaligned systems in this study. Furthermore, we see that this offset is greater for lower misalignment angles and decreases as the misalignment angle increases. This offset also increases as we move farther from the star.

Looking at each panel we also notice small changes in the line of nodes over one orbital period. In all three cases the direction of the tilt seems to oscillate with a period of half a phase. The amplitude of this oscillation also appears to increase with misalignment angle.

Finally, we note that both the tilting angle and line of nodes appears to fluctuate at distances greater than 5 stellar radii. This is likely do to the fact that Equation~\ref{eq:tilt} assumes that the tilted sections are perfectly circular, which is a good approximation close to star but, in the outer discs, the two-armed spiral arms develop as a result of the individual orbits of the particles becoming elliptical. Therefore, a circular fit fails to track the motions of the ensemble of particles around the star.

\subsection{Scale Height}

The scale height (or effective thickness) of the disc is an important quantity sometimes assumed or estimated for modelling discs. Variations in scale heights can have a significant effect on the density structure and the temperature distribution of the disc, which, in turn, will have an effect on the estimation of observables such as emission line profiles and IR continuum levels \citep{car08}. Therefore we investigate the disc scale height in this work. For an isolated Be star, the scale height is defined as
\begin{equation}
H_{\mathrm{th}} \equiv H(r) = c_{s} / \Omega(r),
\label{eq:h}
\end{equation} 
where $\Omega$ is the Keplerian angular velocity at a distance $r$ from the star \citep{lig74}. This relationship is obtained by assuming that the disc is in vertical hydrostatic equilibrium. A derivation of this equation can be found in \citet{car08}. Equation~\ref{eq:h} assumes the presence of a single source of gravitational potential, the Be star. It is therefore important to determine whether or not Equation~\ref{eq:h} is appropriate to estimate the scale height in Be binary systems that are aligned and misaligned. 

In order to extract the scale height throughout the disc, we computed the vertical density structure of the gas at various points in the disc. The vertical density structure of the disc can be described using a Gaussian function
\begin{equation}
\rho(z) \propto e^{-z^2/2H_{\mathrm{sim}}^2},
\label{eq:gauss1}
\end{equation}
where $H_{\mathrm{sim}}$ is the scale height of the disc. Equation~\ref{eq:gauss1}, however, assumes a Gaussian centred at $z=0$, which is a good assumption for isolated Be star systems or aligned systems but not for misaligned systems where we have a warped disc. We therefore modified Equation~\ref{eq:gauss1} to include a second fitting parameter, $\lambda$, which represents how shifted the center of mass is at this particular location compared to $z=0$:
\begin{equation}
\rho(z) \propto e^{-(z-\lambda)^2/2H_{\mathrm{sim}}^2}.
\label{eq:gauss2}
\end{equation}

We then use this relation to fit the vertical density structure of the disc at various locations in order to calculate the scale height for our simulated discs.

\begin{figure}
\includegraphics[width = 0.9\columnwidth]{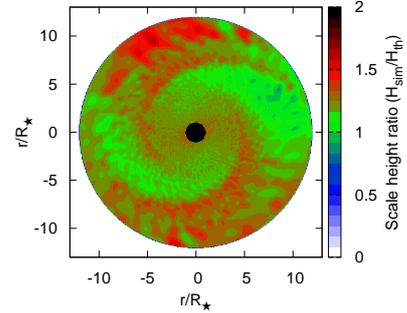}
\caption{Ratio between the scale height obtained from simulations ($H_{\mathrm{sim}}$) and theoretical scale height ($H_{\mathrm{th}}$) computed from Equation~\ref{eq:h} for short period systems with $\alpha_\mathrm{SS} = 0.5$ and misalignment angles of 30$\degr$. The disc is truncated at 12 stellar radii in this Figure.}
\label{fig:dh12}
\end{figure}

Figure~\ref{fig:dh12} shows the ratio between the scale height obtained through simulations ($H_{\mathrm{sim}}$) and the theoretical scale height ($H_{\mathrm{th}}$) computed from Equation~\ref{eq:h}, for short period systems, $\alpha_\mathrm{SS} = 0.5$, and with misalignment angles of 30$\degr$.

Overall we see that Equation~\ref{eq:h} is relatively good at estimating the scale height. However, deviations between simulations and theory can be seen in regions where density waves typically appear in Be-binary systems \citep[see figure 4 of Paper I and figures 10 and 11 of][]{oka02}. This indicates that Equation~\ref{eq:h} might not adequately describe the scale height in the regions of the disc where these density enhancements appear. Note that these results are consistent for all misalignment angles. 

\section{Discussion and Conclusion}

In this investigation we use numerical simulations to study the behaviour of Be star discs in aligned and misaligned binary systems. Simulations were obtained using a smoothed-particle hydrodynamics (SPH) code following previous methods developed and described in the literature \citep{ben90a,ben90b,oka02}. In our simulations, we varied the degree of misalignment between the plane of the orbit and the plane of the disc, the viscosity parameter, $\alpha_\mathrm{SS}$, and the orbital period of the system, $P_{orb}$.

First we studied the azimuthally averaged surface density, $\left < \Sigma(r)\right > _{\phi}$, by investigating the density fall-off rate in the inner and outer parts of the disc ($m$ and $n$ respectively), and the truncation radius ($R_t$). We find, as reported in Paper I for coplanar orbits, that material tends to build up in the innermost part of the disc, a phenomenon called the accumulation effect, for our misaligned systems with the density fall-off values of $m$ reaching values below 2 for most systems. Furthermore we find that the degree of misalignment does have an effect on the magnitude of the accumulation, with a greater accumulation (lower $m$) in aligned systems and with less accumulation in misaligned systems. This difference between aligned and misaligned systems is found to also increase with viscosity. Only for discs with the highest viscosity ($\alpha_\mathrm{SS} = 1.0$) and the longest orbital period ($P_{orb}$ = 60 days) do we find little evidence of accumulation. As reported in Paper~I, both $\alpha_\mathrm{SS}$ and $P_{orb}$ also have an influence on the accumulation found in the inner disc. Similarly we find that $R_t$ also has a dependence on the misalignment angle: discs in aligned systems are found to be truncated closer to the star compared to discs in misaligned systems, with the truncation farther from the primary as the misalignment increases. This is in agreement with the findings \citet{lub15} that discs in misaligned systems have a greater radial extent than their counterparts in aligned systems. We also find that $\alpha_\mathrm{SS}$ and $P_{orb}$ have similar effects on the truncation, that is, the truncation is farther from the primary as both $\alpha_\mathrm{SS}$ and $P_{orb}$ increase. This is also in agreement with the results of Paper~I. Finally, $n$ is found to vary with the misalignment of the system as well. Larger misalignment angles result in denser outer discs, i.e. smaller $n$.

We also studied the warping of the disc by computing the tilting angle and line of nodes of ring segments in the inner part of the disc. We find that the tilt of the disc increases steadily as we move farther from the star, reaching a peak around 4-5 stellar radii. The maximum tilt is found to be greater at a misalignment angle of 45$\degr$ compared to 30$\degr$ and 60$\degr$. The direction of the line of noes is found to vary with both distance from the primary star and with orbital phase. We also find that an offset of about 90$\degr$ between the line of nodes of the disc and of the orbit. This offset is found to decrease when the misalignment angle of the system increases.

Our findings about disc tilting are interesting as they may help to understand the variations in the polarimetric position angle observed in many Be stars. Linear polarization in Be stars is due primarily to the scattering of starlight by free electrons in the innermost region of the disc \citep{coy69, zel72} and is used to measure the position angle of Be stars in the sky. Changes in position angle are due to either variations in mass injection from central star or geometric changes in innermost part of disc. Our results suggests that these variation can also be explained by the presence of a misaligned binary companion. For example, \citet{vin95} found evidence of small cyclical variations, 30$\degr$ or smaller, in the polarimetric position angle of multiple Be stars, which is in agreement with our results. 

Finally we compared the scale heights of our discs with the theoretical scale heights calculated using the standard equation for the case of an isolated Be star \citep{car08}. We find that overall this approximation is relatively good at describing the scale height for all misalignment angles. However, this standard prescription is too simplistic for describing the scale height in regions where density enhancements are present. This should be given careful consideration in future studies.

\citet{mar14} studied Kozai-Lidov oscillations in hydrodynamic discs and showed that for highly misaligned systems this mechanism could be set up. Figure 1 in \citet{fu15} shows that conspicuous disc oscillations can occur for orbital warpings that are large (i.e. 60 and 70$^{\circ}$). During these oscillations the disc can become quite eccentric over time as disc tilting and eccentricity are interchanged. \citet{mar14} suggest that these oscillations may be important for transient discs in Be/Xray systems. Using equation~3 in \citet{fu15} we find that Kozai oscillations would not be expected for our simulations computed to 50 P$_{orb}$. According to our calculations the simulations would need to be followed for approximately 80 P$_{orb}$ in order to see evidence of Kozai oscillations in the inner disc.  Nevertheless, in a future study we plan to study binaries in eccentric orbits and we plan to closely examine the disc shape and variations in inclination to search for the presence of Kozai oscillations.

Given we find that the disc density distribution and the geometry of the disc is altered in misaligned systems, we plan to adapt our codes so that we can compute observables directly from our simulations. Changes in disc density in the inner disc would be seen in observables such as polarization, UBV photometric colours, and spectroscopy, while interferometry, infrared and radio measurements could be used to predict the overall size and density in the disc. Changes in shape and strength of spectral lines could occur, depending on where they are formed in the disc, especially if portions of the disc or central star become obscured along the observers line of sight due to warping or tilting of the disc. As well, in future, we plan to extend this work to investigate the effects on the disc for non-circular binary systems.

\section*{Acknowledgements}
We thank the anonymous referee for comments and suggestions which helped us improve the quality of this paper. This work was made possible by the facilities of the Shared Hierarchical Academic Research Computing Network (SHARCNET:www.sharcnet.ca), Compute/Calcul Canada, and Laboratory of Astroinformatics (IAG/USP, NAT/Unicsul, FAPESP grant No 2009/54006-4). CEJ acknowledges support from NSERC, the Natural Sciences and Engineering Research Council of Canada. DP acknowledges support from FAPESP grant No 2013/16801-2 and CNPq (300235/2017-8). ACC and CEJ thanks UofT-FAPESP-UWO joint researcher program. ACC acknowledges the support from CNPq (grant 307594/2015- 7) and FAPESP (grant 2015/17967-7).


\label{lastpage}
\end{document}